\documentclass[journal,12pt,onecolumn, doublespace]{IEEEtran}
\ifCLASSINFOpdf
\else
\fi
\usepackage{graphicx}
\usepackage{cite}
\usepackage{array}
\usepackage{amsmath}
\usepackage{amssymb}
\usepackage{textpos}
\usepackage{subfigure}
\usepackage{enumerate}
\usepackage{enumitem}
\usepackage{lscape}
\usepackage{float}
\usepackage{pdfsync}
\usepackage{eurosym}
\usepackage{balance}
\usepackage{url}
\usepackage{color}
\usepackage{threeparttable}
\usepackage{multirow}
\usepackage{hyperref}
\usepackage{amsthm}
\usepackage{algorithm}
\usepackage{algorithmic}
\usepackage[latin1]{inputenc}
\usepackage{tikz}
\usetikzlibrary{shapes,arrows}
\usepackage{mathtools}

\newtheorem{remark}{Remark}
\newtheorem{prop1}{Proposition}

\newtheorem{lem1}{Lemma}
\usepackage[font={footnotesize}]{caption}

\renewcommand{\d}{{\rm d}}

\renewcommand{\d}{{\rm d}}

\newcommand{\LOS}{{\mathcal{L}}}
\newcommand{\NLOS}{{\mathcal{N}}}

\DeclarePairedDelimiter{\ceil}{\lceil}{\rceil}
\DeclarePairedDelimiter\floor{\lfloor}{\rfloor}

\begin{document}
\title{Analysis of Cached-Enabled Hybrid Millimter Wave \& Sub-6 GHz Massive MIMO Networks}
\author{
\IEEEauthorblockN{Tong Zhang, Sudip Biswas, Keshav Singh, and Tharmalingam Ratnarajah\vspace{-3.5em}}
\thanks{\hrulefill}

\thanks{The authors are with Institute for Digital Communications, School of Engineering, University of Edinburgh, Edinburgh, UK. Email: \{t.zhang, sudip.biswas, k.singh, t.ratnarajah\}@ed.ac.uk.}
}
\maketitle
\begin{abstract}
This paper focuses on edge caching in mm/$\mu$Wave hybrid wireless networks, in which all mmWave SBSs and $\mu$Wave MBSs are capable of storing contents to alleviate the traffic burden on the backhaul link that connect the BSs and the core network to retrieve the non-cached contents. The main aim of this work is to address the effect of capacity-limited backhaul on the average success probability (ASP) of file delivery and latency. In particular, we consider a more practical mmWave hybrid beamforming in small cells and massive MIMO communication in macro cells. Based on stochastic geometry and a simple retransmission protocol, we derive the association probabilities by which the ASP of file delivery and latency are derived. Taking no caching event as the benchmark, we evaluate these QoS performance metrics under MC and UC placement policies. The theoretical results demonstrate that backhaul capacity indeed has a significant impact on network performance especially under weak backhaul capacity. Besides, we also show the tradeoff among cache size, retransmission attempts, ASP of file delivery, and latency. The interplay shows that cache size and retransmission under different caching placement schemes alleviates the backhaul requirements. Simulation results are present to valid our analysis.
\end{abstract}
\vspace{-1.5em}
\begin{IEEEkeywords}\vspace{-1.5em}
Wireless edge caching, latency, hybrid millimetre-micro wave network, Poison point processes.\vspace{-1.50em}
\end{IEEEkeywords}
\vspace{-1.0em}	
\section{Introduction}\vspace{-0.5em}
The huge mobile traffic in wireless communications, mainly caused by the the mobile video traffic that accounts for the majority of the total mobile data traffic, has brought us needs for 5G and beyond 5G technologies \cite{Cisco_1, andrews2014will}. 
Currently, massive multiple-input multiple-output (MIMO) communication, millimeter-wave communication, and network densification through heterogeneous networks (HetNets) are the most three promising techniques proposed for 5G wireless communication systems. 
However, though the above potential solutions are beneficial for the access links, they do little to alleviate the burden on the backhaul links that connect edge base stations (BSs) to the data center in the core network. Further, their requirement for the existence of expensive backhaul links exaggerates the backhaul congestion issue during peak hours. 
In particular, it is found that only a small percentage (5--10\%) contents (i.e., treated as the popular contents) are repeatedly requested by the majority of users, which results in a substantial amount of the redundant data traffic over networks. Motivated by this, caching the popular contents at the edge nodes closer to users (i.e., wireless edge caching) is proposed as a promising solution to offload the traffic of backhual links and reduce the backhaul cost and latency.
Caching technique includes two different phases. while the first phase is caching placement phase that is conducted during off-peak hours according to the statistics of the users' requests and the main limitation of this phase is the caching capacity, the second is content delivery phase that is performed after the actual requests of the users have been revealed and the main limitation of this phase is the QoS requirements. 

As aforesaid, since the caching capacity at edge nodes is limited and much less than the total amount of popular contents of users' interest, it is necessary to design proper caching placement strategies to make exploit use of caching benefit. However, due to  more dynamic wireless networks than wired networks, implementing caching technique in wireless networks is more challenging than wired networks and wired caching strategies cannot be directly applied to wireless networks. Put another way, the unique transmission features and randomness in cellular networks, e.g., fading channel, limited spectrum, and co-channel interference, are required to take into consideration when designing efficient caching strategies.

Recent studies have focused on the caching design and analysis in various scenarios. Both centralized and decentralized coded caching were studied in a basic model with a shared error-free link to acquire more caching gains by creating more multicast transmission in \cite{maddah2014fundamental} and \cite{maddah2015decentralized}. Futher, by taking into network topology into consideration, the optimal caching placement strategies were designed to minimize the average sum delay for both coded and uncoded scenarios in a simple cache-enabled femtocell networks (i.e., caching helper networks) \cite{shanmugam2013femtocaching}. Futher, the throughput scaling law was studied with the random caching strategy in a simple grid-modelled D2D networks \cite{ji2016fundamental}. However, the network models considered in \cite{maddah2014fundamental,maddah2015decentralized,shanmugam2013femtocaching,ji2016fundamental} did not capture the stochastic natures of channel fading, interference, annd geographic locations of network nodes. In order to take account of realistic cellular networks, some other works focused on caching technique in a stochastic geometric framework. In \cite{chae2016caching}, a probabilistic caching model was applied to a single-tier stochastic wireless caching helper networks and the optimal caching placement was designed in terms of average success probability of delivery for both noise-limited and interference-limited scenarios. In \cite{chae2017content}, caching cooperation was studied in a same caching helper networks and the near-optimal caching placement scheme that maximizes the approximated average success probability of delivery was acquired. Further, optimal caching placement strategy in the $N$-tier HetNets was designed, where the optimal caching probabilities maximize the average success probability of delivery \cite{li2018optimization}. With stochastic geometric framework, D2D caching was investigated in literature such as \cite{chen2017probabilistic}, \cite{wang2017cooperative}. While \cite{chen2017probabilistic} evaluated the throughput gain acquired with different optimal cache-hit and throughput caching placement strategies, \cite{wang2017cooperative} considered a D2D underlaid cellular network in which an optimal cooperative caching placement was studied whose performance of the average success probability (ASP) outperforms other caching placement strategies.

However, on one hand, prior works focused on cache hit event to acquire the optimal caching strategies but they did not take cache miss event into consideration and did not justify the design and insights of caching on backhaul limitations; on the other hand, caching capacity of local BSs can be treated as a new type of resources of wireless networks other than time, frequency, and space. Therefore, the emerging radio-access technologies and wireless network architectures provide edge caching new opportunities to fulfil the common goals of improving the quality of service (QoS) and quality of experience (QoE) for users, which makes it imperative to investigate the performance of these technologies in a co-existed framework. 
In this regard, the work \cite{peng2015backhaul} evaluated the impact of backhaul delay and proposed an optimal caching strategy with respect to average download delay while \cite{yang2016analysis} considered backhaul effect on throughput and delay analysis in multi-tier HetNets. Moreover, the backhaul effect were also taken into consideration in literature \cite{song2018minimum, fan2018energy}. While \cite{song2018minimum} aimed to find out the tradeoff between optimal cache size and backhaul requirement, the work \cite{fan2018energy} aimed to analyze the impact of backhaul on the energy efficiency in HetNets. From another research direction, research work has focused on caching in mmWave networks and mmWave/$\mu$Wave hybrid networks in literature \cite{zhu2018performance, giatsoglou2017d2d, biswas2018analysis,zhu2018content} due to the problem of $\mu$Wave spectrum crunch. However, none of the above work have studied the cache-enabled hybrid HetNets with limited backhaul transmission and considered a relatively practical mmWave hybrid beamforming together with massive MIMO.   
 
In this paper, we focus on edge caching in a backhaul limited mm/$\mu$Wave hybrid network assisted with massive MIMO, which has been understood yet, especially considering the mmwave hybrid beamforming. On one hand, mmWave hyrbid beamforming is motivated by the fact that the cost and power hungry for large-scale antenna arrays at mmWave bands; on the other hand, the backhaul implementation cost is very expensive, especially for the large capacity backhaul. Therefore, it is necessary to investigate the what network design parameters can help alleviate the backhaul capacity requirement and how backhaul impact works on different performance metrics.  
Our work contributions are summarized as follows:
\begin{itemize}
\item We consider cache-enabled hybrid HetNets, where the locations of nodes are modelled as homogeneous Poison point processes (PPPs). In particular, small BSs (SBSs) aided with mmWave hybrid beamforming operated at mmWave frequencies and macro BSs (MBSs) aided with massive MIMO operated at $\mu$Wave frequencies are equipped with finite cache size to store some popular contents. Moreover, we also consider limited backhaul links between BSs and the gateways, which has not been studied in the existing literature.
\item We first derive the association probabilities by which the probability of the typical user is associated with different BSs is characterized. Then we derive the PDF of distance between the serving BS and the typical user. 
\item Considering mmWave and $\mu$Wave transmission, we derive retransmission based ASP of file delivery, latency, and backhaul load per unit area based on stochastic geometry.
\item Taking no caching as the benchmark, we numerically analyze the performance of ASP of file delivery, latency, and backhaul load per unit area under different caching strategies with respect to many significant network design parameters, such as cache size, antenna number, backhaul capacity, blockage density, target data rate, blockage density, the number of retransmission attempt, and content popularity. We demonstrate that weak and strong backhaul have different impacts on different caching strategies due to association probabilities, e.g., when the backhaul capacity is huge, UC performs worse than no caching case. Besides, we also evaluate the tradeoff between cache size and backhaul capacity. Moreover, we confirm that retransmission is a good solution to improve QoS by increasing retransmission attempts but we also show the tradeoff between ASP of file delivery and latency.  
\end{itemize}
\section{System Model}
In this section, we introduce the network topology, the caching model, the cache-enabled content access protocol, the partial probabilistic caching placement schemes, and the user association policy. The main notations used in this paper are summarized in Table \ref{Notation Summary}.
\begin{table*}
\renewcommand{\arraystretch}{0.75}
\centering
\caption{Notation Summary}
\label{Notation Summary}
\begin{tabular}{|l|l|}
\hline
{Notations} & {Physical meaning} \\ \hline
{$\Phi_\mu$, $\Phi_m$, $\Phi_u$} & {PPP distributed locations of $\mu$Wave MBSs, mmWave SBSs, and UEs}\\ \hline
{$\lambda_\mu$, $\lambda_m$, $\lambda_u$, $\lambda_g$} & {Spatial densities of $\mu$Wave MBSs, mmWave SBSs, UEs, and gateways} \\ \hline
{$n^\mu_t$, $n^m_t$} & {Number of transmit antennas at each $\mu$Wave MBS and mmWave SBS}\\ \hline
{$n^\mu_r$, $n^m_r$} & {Number of receive antennas at each UEs} \\ \hline
{$\mathrm{P}_\mu$, $\mathrm{P}_m$} & {Transmit power of each $\mu$Wave MBS and mmWave SBS} \\ \hline
{$\mathcal{F}$ \emph{i.e.,} $|\mathcal{F}| = F$} & {The limited file set with $F$ files} \\ \hline
{$f_i$ with $\forall i \in \mathcal{F}$ } & {The probability for requesting the $i$th file} \\ \hline
{$C_\mu$, $C_m$} & {Cache sizes of each $\mu$Wave MBS and mmWave SBS} \\ \hline
{$S$} & {The number of bits per file} \\ \hline
{$C_b$} & {The backhaul capacity per BS (either mm- or $\mu$-Wave)} \\ \hline
{$x^\LOS_{m_i}$, $x^\NLOS_{m_i}$, $y^\LOS_{m_i}$, $y^\NLOS_{m_i}$ } & {Locations of the associated mmWave SBSs with LOS and NLOS transmission for}\\ 
{}&{the $i$th file } \\ \hline
{$x_{\mu_i}$, $y_{\mu_i}$} & {Locations of the associated cache hit and cache miss $\mu$Wave MBSs for the $i$th file} \\ \hline
{$r_{x^\LOS_{m_i},0}$, $r_{x^\NLOS_{m_i},0}$, $r_{y^\LOS_{m_i},0}$, $r_{y^\NLOS_{m_i},0}$} & {Distances between the associated mmWave SBSs with LOS and NLOS}\\ 
{}&{transmissions and the typical user} \\ \hline
{$r_{x_{\mu_i},0}$, $r_{y_{\mu_i},0}$} & {Distances between the cache hit and cache miss $\mu$Wave MBSs and the typical user} \\ \hline
{$\beta$} & {Blockage density} \\ \hline
{$B_m$, $B_\mu$} & {Biased factors} \\ \hline
{$p_\LOS(\cdot)$, $p_\NLOS(\cdot)$} & {The LOS and NLOS probabilities of the channel link} \\ \hline
{$\mathcal{U}_{\cdot}$} & {The set of users that can be served by a BS} \\ \hline
{$U_{j}$ with $j \in \{m, \mu\}$} & {The number of users served by the associated mmWave or $\mu$Wave BS} \\ \hline
{$n_{\rm RF}$} & {RF chain} \\ \hline
{$N_{\cdot}$} & {The number of users associated with the associated BS} \\ \hline
{$\eta_{j}$ with $j \in \{\LOS, \NLOS\}$} & {The number of paths} \\ \hline
{$\phi$, $\theta$} & {AOA and AOD} \\ \hline
{$ \mathcal{G}$} & {Channel coefficient of $\mu$Wave communication} \\ \hline
{$\mathcal{X}$} & {Channel coefficient of mmWave communication} \\ \hline
\end{tabular}\vspace{-2.5em}
\end{table*}
\vspace{-1.5em}
\subsection{Network Architecture}
We consider the downlink of a two-tier cache-enabled hybrid wireless heterogeneous network, where massive MIMO-aided macro BSs ($\mu$Wave MBSs) operated at sub-6GHz $\mu$Wave spectrum are overlaid with successively denser small BSs (mmWave SBSs) operated at mmWave spectrum.
By utilizing the stochastic geometric framework, SBSs and MBSs are deployed in a 2-D Euclidean plane $\mathbb{R}^2$ based on mutually independent homogeneous Poisson point processes (HPPPs) $\Phi_m$ and $\Phi_\mu$ with densities $\lambda_m$ and $\lambda_\mu$, respectively. Accordingly, all users are also distributed as another PPP $\Phi_u$ with density $\lambda_u$. Particularly, in practical system there are more users than BSs, thus we assume $\lambda_u > \lambda_m > \lambda_\mu$ in this work. {Since mmWave and $\mu$Wave transmissions occur in different frequency bands, both transmissions can be considered to be orthogonal to each other with different transmitting as well as receiving antenna interfaces.} Put another way, the set of BSs or users belonging to a certain network (small cell network (SCN) or macro cell network (MCN)) operate in the same spectrum (mm or $\mu$Wave), it does not interfere with the set of BSs or users from the other network. Further, all MBSs and SBSs are equipped with $n^\mu_t$ and $n^m_t$ antennas with transmit power $\mathrm{P}_\mu$ and $\mathrm{P}_m$, respectively. Considering two different transmissions in the work, the user equipment (UEs) are assumed to be equipped with two sets of RF chains with antennas $n^\mu_r$ and $n^m_r$ to independently received $\mu$Wave and mmWave signals, respectively. 
\begin{remark}
We assume $n^m_r > 1$ and $n^\mu_r = 1$. This is due to the intrinsic relation between wavelength of signals and antenna separation, whereby the wavelength of $\mu$Wave signals is much larger than mmWave signals and hence, much larger separation is required between antennas for $\mu$Wave to avoid correlation and coupling. As a result, accommodating more than one antenna at small devices, like mobile phones operating in the $\mu$Wave spectrum may not be feasible.
\end{remark}
Compared to closed access, this work considers open access, which a user is allowed to be asscosiated with any tier's BSs to provide best-case coverage. For analytical tractability, instead of considering the exact average biased received signal power, we consider a cell association criterion based on least biased path loss with a bias factor that includes the average channel gain and all other effects to control the cell range. This enables the cell association to be tuned for balancing the cell load or other purposes. 
When the bias factor is greater than one, it extends the cell range. Otherwise, it shrinks the cell range. Due to mmWave short propagation distance, it is reasonable to adjust bias factor smaller than one. Based on the least biased path loss association criterion, the user will be served by either a mmWave SBS or a $\mu$Wave MBS with differennt association probabilities. Due to Slivnyak's theorem that ensures the statistics measured at a random point of a PPP is the same as those measured at the origin \cite{7502130}. Therefore, the analysis hereinafter is performed for the typical user located at the origin, denoted by $0$. Besides, it is worth noting that the propagation between the mmWave SBS and the typical user is via a fully connected hybrid precoder that combines radio frequency (RF) and baseband (BB) precoding. Due to the sparsity of mmWave channels, we assume that all scattering happens in the azimuth plane. Therefore, the uniform linear array (ULA) is justifiably assumed to employ at each mmWave BS and UE with size of $n^m_t$ and $n^m_r$, respectively. In contrast, the propagation between the $\mu$Wave MBS and the typical user is via massive MIMO where we do not consider any training in the forward link and therefore, the users do not have nay channel knowledge. In particular, the pilot contamination problem is mainly due to uplink training with non-orthogonal pilots. Since the main work focuses on analysis of caching placement in backhaul-limited hybrid network, the problem of pilot contamination is avoided and not considered by assuming the assignment of pilot sequences to the users who are associated with MBSs is orthogonal. In this regard, we consider that  the number of users associated to MCN is less than the available pilot length.
\vspace{-1.5em}
\subsection{Caching model}
It has been observed that people are always interested in the most popular multimedia contents, where only a small portion of the contents are frequently accessed by the majority of users. This work assumes that the finite file set $\mathcal{F}$ consisting of $F$ multimedia contents on the content server, where all BSs can get access to the content server to retrieve the cache miss contents via the capacity-limited backhaul links. In order to avoid backhaul burden caused by redundant transmissions of the repeated requests during peak hours, caching is implemented at all BSs (including mmWave and $\mu$Wave BSs) but with different cache sizes $C_m$ and $C_\mu$, respectively, such that $C_m < C_\mu$. For clarity, all files in the file set $\mathcal{F}$ are of equal size, denoted by $S$ bits. This assumption is justifiable due to the fact that each file can be divided into multiple chunks of equal size or different sizes. Hereinafter, for notational simplicity, we use file index  to denote each file, namely $\mathcal{F} = \{ 1, 2 , \cdots, F\}$. Then, each mmWave and $\mu$Wave BS can cache up to $C_m \times S$ bits and $C_\mu \times S$ bits, respectively. Further, each user independently and identically requests the $i$th file from the file set $\mathcal{F}$ according to the Zipf distribution given by $f_i = (1/i^\upsilon)/(\sum_{j=1}^{F}(1/j^\upsilon)$, where a skewness parameter $\upsilon$ controls the skewness of the content popularity distribution. The content popularity tends to be uniform distribution when $\upsilon$ goes to zero. However, even though Zipf distribution is commonly utilized in \cite{7880694, 7150324}, especially for videos and web pages, the analysis of this work can also be applied to other content popularity distribution and it is expected to exhibit similar analytical results. 
\vspace{-1.0em}
\subsection{Cache-enabled Push and Delivery Content Access Protocol}\vspace{-0.5em}
Based on the client/server (C/S) structure, all BSs are connected with a default gateway (or central controller) via a capacity-limited wired backhaul solution while the high-capacity wired backhaul solution is assumed for the connection between a gateway and content server, which supports relatively highly reliable transmission. Then this work only considers the effect of backhaul from the connection between the BSs and the centrol controller while neglecting the effect of the backhaul connection between the central controller and the content server. Particularly, the limited backhaul capacity is equally allocated to all BSs, the backhaul capacity of each BS, denoted by $C_b$, is the function of BS density, given as \cite{bacstug2015cache, 7511288}\vspace{-1.0em}
\begingroup\makeatletter\def\f@size{10}\check@mathfonts
\def\maketag@@@#1{\hbox{\m@th\small\normalfont#1}}
\begin{equation}
C_b=\frac{c_1}{\lambda_m + \lambda_\mu} + c_2\,,\label{backhaul capacity}\vspace{-0.50em}
\end{equation}
where $c_1 > 0$, $c_2 > 0$ are arbitrary coefficients with regard to the capacity of backhaul links. The more the number of BSs included in the network, the less capacity of each BS it is.

During peak hours, the users' requests are collected to estimate the content popularity distribution by means of some estimation technologies. For analytical tractability, hereinafter we assume that the popularity of the files is perfectly known and stationary. This assumption is perhaps over simplistic, but we leave the investigation of unknown and time-varying popularity for future work. In particular, for given the time-varying content popularity, the seek of new caching placement schemes and the analysis incorporated with estimation error should be required although it is not addressed in this paper. During off-peak hours, the network traffic load is relatively low and cache placement phase is implemented according to the content popularity distribution and some specific proactive caching placement policies. By pushing some desired contents to  the local caches of the BSs via broadcasting, the aim is to further alleviate the network traffic burden in the content delivery phase during peak hours. In particular, all the cache-enabled BSs proactively fetch the same copy of the contents up to the full cache sizes through some certain caching placement schemes that will be given in details in the following subsection.

When a user requests a content from the file set $\mathcal{F}$, it initially checks if the requested content is available in the local caches of the associated BS. If the requested content is cached at the associated BS, then it directly serves the requesting user without the need of backhaul links. Otherwise, the requested content should be retrieved initially by BSs from the content server in the core network via capacity-limited bakchaul links  modelled by \eqref{backhaul capacity}, then forwarded to the requesting user via wireless access links. As mentioned above, considering the typical user requests the $i$th file, there are total six possible content access and association events, including both cache hit and cache miss scenarios described as follows. In view of the fact that mmWave signals are sensitive to blockages in the network, such as trees, concrete buildings, and so forth, this work considers two different transmission conditions \emph{i.e.,} LOS and NLOS transmissions with different path loss coefficients. This content is provided in detail in the association subsection.

\textbf{Scenario 1}: The typical user associated with a mmWave BS located at $x^{\LOS}_{m_i}$ that has the requested file in its local caches is served in LOS transmission. The distance between the typical user and the associated BS is denoted by $r_{x_{m_i}^\LOS,0}$.

\textbf{Scenario 2}: The typical user associated with a mmWave BS located at $x^{\NLOS}_{m_i}$ that has the requested file in its local caches is served in NLOS transmission. The distance between the typical user and the associated BS is denoted by $r_{x_{m_i}^\NLOS,0}$.

\textbf{Scenario 3}: The typical user associated with a mmWave BS located at $y^{\LOS}_{m_i}$ that has not the requested file in its local caches is served in LOS transmission. The requested file is forwarded to the typical user via backhaul link and access link in order. The distance between the typical user and the associated BS is denoted by $r_{y_{m_i}^\LOS,0}$.

\textbf{Scenario 4}: The typical user associated with a mmWave BS located at $y^{\NLOS}_m$ that has not the requested file in its local caches is served in NLOS transmission. The requested file is forwarded to the typical user via backhaul link and access link in order. The distance between the typical user and the associated BS is denoted by $r_{y_{m_i}^\NLOS,0}$.

\textbf{Scenario 5}: The typical user is associated with a $\mu$Wave BS located at $x_{\mu_i}$ that has the requested file in its local caches. The distance between the typical user and the associated BS is denoted by $r_{x_{\mu_i},0}$.

\textbf{Scenario 6}: The typical user is associated with a $\mu$Wave BS located at $y_{\mu_i}$ that has not the requested file in its local caches. The requested file is forwarded to the typical user via backhaul link and access link in order. The distance between the typical user and the associated BS is denoted by $r_{y_{\mu_i},0}$.

The Scenario 1 -- 4 are for the typical user associated with SCN while Scenario 5 -- 6 are for the typical user associated with MCN. In particular, the work takes no caching scenario as the benchmark where all BSs are not able to cache any files. We will give the association probability of the typical user associated with the aforesaid event in the association subsection. 
\subsection{Caching placement schemes}
As for proactive caching, the content placement is usually conducted during off-peak traffic periods. In this phase, the network prefetches the content to the storage by means of some caching placement strategies to decide which file should be cached in which BSs. Different from wired networks with fixed and known network topology, highly dynamic  wireless network topology makes it impossible to known as a prior that which user will associate with which BS due to undetermined user locations. In order to deal with this problem, this work utilizes the probabilistic content caching policy rather than the deterministic caching policy considered in wired networks, where the contents are independently and randomly cached with given probabilities in a distributed manner, so it can be applicable even to complex networks with high flexibility. 
 
This work applies three probabilistic caching placement schemes -- uniform caching (UC), caching $M$ most popular contents (MC), and random caching (RC) -- that are commonly utilized in most existing work. In particular, we consider all the BSs in a same tier with the same caching probabilities and each BS caches contents with the given probabilities independently of other BSs. 
For clarity, we define the caching probabilities that $\mu$Wave and mmWave BSs caching the contents as $P_j = \{p_{j_1}, p_{j_2}, p_{j_3}, \cdots, p_{j_i}, p_{j_{i+1}}, \cdots, p_{j_{F}} \}$ with subscript $j \in \{m, \mu\}$ denoting either mmWave SCN or $\mu$Wave MCN, respectively. Based on thinning theorem, the thinned PPP $\Phi_{j_i}$ consisting of the BSs storing the $i$th file is characterized by the density $\lambda_j p_{j_i}$.

As described in \cite{7502130, 7995138}, UC, MC, and RC are given in a mathematical expression as follows.

\textbf{UC}: The contents in the file set $\mathcal{F}$ are uniformly selected to be cached in the local caches with caching probabilities given as $p_{j_i} = {C_j}/ F $ with $j \in \{m,\mu\}$ and $i \in [1, F] \triangleq \{1, 2,  \cdots, F\}$.

\textbf{MC}: The first ${C_j}$ contents in the file set $\mathcal{F}$ are certainly selected to be cached in the local caches with caching probabilities given as \vspace{-1.0em}
\begin{align}
p_{j_i}= \Bigg\{
\begin{array}{ll}
1, & i\in [1, {C_j}]\\
0, & i\in[{C_j}+1, F]
\end{array}\,,
\end{align}
with $j\in\{m,\mu\} \text{ and } i\in [1,F]$.

\textbf{RC}: The contents in the file set $\mathcal{F}$ are randomly selected to be cached in the local caches with caching probabilities given as
$\forall i\in[1,F]\quad 0\le p_{j_i} \le 1 \text{ s.t.} \sum_{i=1}^{F} p_{j_i} \le {C_j}$ with $j \in \{m,\mu\}$. In fact, it is vague to use the term random caching since we have not defined the distribution to generate the caching probabilities. For simplicity, this work considers that each caching probability $p_{j_i}$ uniformly chooses a value between $0$ and $1$. In particular, the summation of all $F$ caching probabilities has the mean of $\frac{F}{2}$. However, when the cache size is too small such that $C_j < \frac{F}{2}$, it will never generate a valid realisation of such a random caching probability by all means. In order to deal with it, we introduce a scaler $(\frac{F}{2 C_j})^{-1}$ to make the mean approximately equal to $C_j$. In this manner, it is reasonable to expect that random caching is slightly better than uniform caching. In fact, the performance of random caching is highly related to this generating distribution. 

Finally, no caching is defined as $p_{j_i} = 0, \, \forall i \in [1, F]$ with $j \in \{m, \mu\}$, which is considered as the benchmark.
\vspace{-1.50em}
\subsection{Association probability}\vspace{-0.50em}
Unlike other existing work that considers the users only associated with BSs caching the requested files to find the optimal caching placement, this work considers both cache hit and cache miss association scenarios to figure out the backhaul effects. Besides, due to mmWave LOS/NLOS transmissions, mmWave SBS coverage areas are no longer the typical weighted Voronoi cell since a user can associate with a far away BS with LOS transmission rather than a closest BS with NLOS transmission. Thus, least distance association criterion is not suitable any more. For simplicity, we consider  least biased path loss as the user association strategy instead of maximum biased received signal power. The  least biased path loss for mmWave and $\mu$Wave BSs are given by $B_m r^{-\alpha_j}$ with $j \in \{\LOS, \NLOS\}$ and $B_\mu r^{-\alpha_\mu}$, respectivley. The bias factor is to control the cell range. Usually they are set as 1's in \cite{6287527}. However, due to mmWave short propagation distance, we slightly shrink the mmWave SBS coverage area by setting $B_m$ less than 1 while keeping $B_\mu$ as 1. 

This work adopts a two state statistical blockage model for each link as in \cite{7448962}, such that the probability of the link to be LOS or NLOS state is a function of the distance between the typical user and the serving mmWave SBS. Assume the distance between them is $r$, the probability of the link is LOS state is $p_\LOS(r) = e^{-\beta r}$ and NLOS state is $p_\NLOS(r) = 1 - e^{-\beta r}$ where $\beta$ is the blockage density. Based on the thinning theorem, the PPP $\Phi_m$ is further thinned as $\Phi^\LOS_m$ and $\Phi_m^\NLOS$ in terms of LOS and NLOS states with density $\lambda_m p_\LOS(r)$ and $\lambda_m p_\NLOS(r)$, respectively.  

As described in the above section, we have defined 6 association events. Now the following Lemma \ref{association mmWave} and \ref{association muwave} gives the 6 association probabilities for the typical user associated with SCN and MCN, respectively.

\begin{lem1}\label{association mmWave}
The probabilities that the typical user requesting the $i$th file is associated with cache hit mmWave SBSs located at $x_{m_i}^\LOS$ and $x_{m_i}^\NLOS$ with LOS and NLOS transmissions are given by\vspace{-0.5em}
\begin{align}
p_{x_{m_i}^\LOS} =&\int_{0}^{\infty} \!\!\!\exp\Big(-\pi \lambda_\mu (\frac{B_\mu R^{\alpha_\LOS}}{B_m})^{\frac{2}{\alpha_\mu}}\Big)\exp\Big(- 2 \pi \lambda_m Z(R) \Big) \nonumber \\
&\times\exp\Big( -2 \pi \lambda_m \hat{Z}(R^{\frac{\alpha_{\LOS}}{\alpha_{\NLOS}}}) \Big) 2 \pi \lambda_m p_{m_i} R e^{-\beta R} \textup{d} R\,,
\end{align}
\begin{align}
p_{x_{m_i}^{\NLOS}}=&\int_{0}^{\infty}\!\!\!\!\!\!\exp\Big(-\pi \lambda_\mu (\frac{B_\mu R^{\alpha_\NLOS}}{B_m})^{\frac{2}{\alpha_\mu}}\Big)\exp\Big( \!\!- \!2 \pi \lambda_m Z(R^{\frac{\alpha_{\NLOS}}{\alpha_{\LOS}}})\Big)\nonumber \\
& \times\exp\Big(\!-\!2\pi \lambda_m \hat{Z}(R)\Big) 2 \pi \lambda_m p_{m_i} \Big(R \!\!-\!\! R e^{-\beta R}\Big)\textup{d} R\,,
\end{align}
and the probabilities that the typical user requesting the $i$th file is associated with cache miss mmWave SBSs located at $y_{m_i}^\LOS$ and $y_{m_i}^\NLOS$ with LOS and NLOS transmissions are given by\vspace{-0.5em}
\begin{align}
p_{y_{m_i}^\LOS}=&\int_{0}^{\infty} \!\!\!\exp\Big(-\pi \lambda_\mu (\frac{B_\mu R^{\alpha_\LOS}}{B_m})^{\frac{2}{\alpha_\mu}}\Big)\exp\Big(- 2 \pi \lambda_m Z(R) \Big) \nonumber\\
&\times\exp\Big( -2 \pi \lambda_m \hat{Z}(R^{\frac{\alpha_{\LOS}}{\alpha_{\NLOS}}}) \Big) 2 \pi \lambda_m (1-p_{m_i}) R e^{-\beta R} \textup{d} R\,,
\end{align}
\begin{align}
p_{y_{m_i}^{\NLOS}}=&\int_{0}^{\infty}\!\!\!\!\!\!\exp\Big(-\pi \lambda_\mu (\frac{B_\mu R^{\alpha_\NLOS}}{B_m})^{\frac{2}{\alpha_\mu}}\Big)\exp\Big( \!\!- \!2 \pi \lambda_m Z(R^{\frac{\alpha_{\NLOS}}{\alpha_{\LOS}}})\Big) \nonumber \\
&\times\exp\Big(\!-\!2\pi \lambda_m \hat{Z}(R)\Big) 2 \pi \lambda_m (1\!-\!p_{m_i}) \Big(R \!\!-\!\! R e^{-\beta R}\Big)\textup{d} R\,,
\end{align}
where $Z(R) = - \frac{1}{\beta} R e^{-\beta R} - \frac{1}{\beta^2} (e^{-\beta R} - 1 )$ and $\hat Z(R) = \frac{R^2}{2} - Z(R)$.
\end{lem1}
\begin{proof}
The proof of the Lemma is given in the Appendix \ref{Appears_A}. For simplicity, only the proof of the probability of the typical user associated with the mmWave SBS caching the requested file with LOS transmission is given in the Appendix \ref{Appears_A}. Other association probabilities can be proofed in a similar way.
\end{proof}
\vspace{-1.0em}
\begin{lem1}\label{association muwave}
The probability that the typical user requesting the $i$th file is associated with cache hit $\mu$Wave MBS located at $x_{\mu_i}$ is given by\vspace{-0.5em}
\begin{align}
p_{x_{\mu_i}}=& \int_{0}^{\infty}\!\!\!  \exp\Big[-\pi\lambda_\mu R^2\Big] \exp\Big[-2\pi\lambda_m Z\Big((\frac{B_m  R^{\alpha_\mu}}{B_\mu })^{\frac{1}{\alpha_\LOS}}\Big)\Big]\nonumber \\
&\times \exp\Big[-2\pi\lambda_m \hat{Z}\Big((\frac{B_m  R^{\alpha_\mu}}{B_\mu })^{\frac{1}{\alpha_\NLOS}}\Big)\Big] 2 \pi \lambda_\mu p_{\mu_i} R\textup{d} R\,,
\end{align}
and the probability that the typical user requesting the $i$th file is associated with cache miss $\mu$Wave MBS located at $y_{\mu_i}$ is given by\vspace{-0.50em}
\begin{align}
p_{y_{\mu_i}} =&\int_{0}^{\infty}  \exp\Big[-\pi\lambda_\mu R^2\Big] \exp\Big[-2\pi\lambda_m Z\Big((\frac{B_m  R^{\alpha_\mu}}{B_\mu })^{\frac{1}{\alpha_\LOS}}\Big)\Big]\nonumber \\
&\times \exp\Big[-2\pi\lambda_m \hat{Z}\Big((\frac{B_m  R^{\alpha_\mu}}{B_\mu })^{\frac{1}{\alpha_\NLOS}}\Big)\Big]2 \pi \lambda_\mu (1-p_{\mu_i}) R \textup{d} R\,,
\end{align}
where all parameters have already defined in Lemma \ref{association mmWave}.
\end{lem1}
\begin{proof}
The proof of the Lemma is similar to the Lemma \ref{association mmWave}.
\end{proof}
\vspace{-1.0em}
\subsection{Average Number of Users}\vspace{-0.5em}
Each mmWave SBS and $\mu$Wave MBS serves multiple users simultaneously with equal power allocation. Consequently, the link capacity of each user is reduced by a fraction of the number of users served simultaneously. If the numbers of users simultaneously served by a mmWave and a $\mu$Wave BS located at $x_m$ and $x_\mu$ are denoted as $N_{x_m}$ and $N_{x_\mu}$, respectively. Since the association coverage areas are different from a distance-based Poisson-Voronoi cells, it is complicated to compute the exact cell distribution. Consequently, this work considers the average number of users following the same assumption in \cite{7502130}, where the average number of users are given by assuming the same mean cell areas as that of the Poisson-Voronoi cell areas.  
The average number of users associated with the tagged mmWave and $\mu$Wave BSs are given by \cite{7110547 ,6497002} \vspace{-0.5em}
\begin{align}
\begin{array}{ll}
N_{x_m} = 1 + 1.28 \frac{\lambda_u p_{am}}{\lambda_m}\,,\;\;N_{x_\mu}  = 1 + 1.28 \frac{\lambda_u p_{a\mu}}{\lambda_\mu}
\end{array}\,,
\end{align}
where $p_{am}$ and $p_{a\mu}$ are the association probability of the typical user associated with SCN and MCN, respectively, such that $p_{am} = p_{x_{m}^\LOS} + p_{x_m^\NLOS} + p_{y_m^\LOS} + p_{y_m^\NLOS}$ and $p_{a\mu} = p_{x_\mu} + p_{y_\mu}$. In particular, it is worth noting that $p_{am}$ and $p_{a\mu}$ are not the function of caching probabilities since the total number of users associated with SCN and MCN includes all cache hit and cache miss users, by which the  caching probabilities are averaged out.

Accordingly, the numbers of associated users of the other mmWave and $\mu$Wave BSs except the tagged mmWave and $\mu$Wave BS are given as\vspace{-0.5em}
\begin{align}
\begin{array}{ll}
N_{\bar x_m} = \frac{\lambda_u p_{am}}{\lambda_m}\,,\;\;N_{\bar x_\mu}  = \frac{\lambda_u p_{a\mu}}{\lambda_\mu}
\end{array}\,,
\end{align}
where $\lambda_m p_{am}$ and $\lambda_\mu p_{a\mu}$ are the number of users associated with SCN and MCN, respectively. 

However, in practice, due to the finite number of RF chains (antennas), the associated users to serve should not be more than the available number of RF chains (antennas) in one time/frequency resource block. Assume the set of the served users of each mmWave located at $x_{m}$ and $\mu$Wave BS located at $x_\mu$ are denoted by $\mathcal{U}_{x_m}$ and $\mathcal{U}_{x_\mu}$, respectively. Accordingly, the cardinalities of the sets $\mathcal{U}_{x_m}$ and $\mathcal{U}_{x_\mu}$ are given by $U_{x_m}$ and $U_{x_\mu}$, respectively. Unlike mmWave hybrid beamforming, this paper applied the fully-digital baseband processing  to massive MIMO where each RF chain per antenna is applied. However, this approach is impractical for mmWave BSs equipped with much larger antenna arrays than massive MIMO. Therefore, hybrid beamforming\footnote{The next section provides the more details.} is implemented and the number of RF chains is not less than the number of antennas due to hardware complexity, power consumption, and cost. Consider the mmWave RF chains is $n_{\text{RF}}$, the total number of served UEs of a tagged $\mu$Wave and tagged mmWave BS are $U_{x_\mu} = \min(n^t_\mu, N_{x_\mu})$ and $U_{x_m} = \min(n_{\text{RF}}, N_{x_{m}})$, respectively. The total number of users of any other $\mu$Wave BS and mmWave BS except the tagged $\mu$Wave and mmWave BS are $U_{\bar x_\mu} = \min(n^t_\mu, N_{\bar x_\mu})$ and $U_{\bar x_m} = \min(n_{\text{RF}}, N_{\bar x_m})$, respectively. For notational simplicity, hereinafter the average number of users of the tagged mmWave BS and other mmWave BS except the tagged mmWave BS are expressed as $U_m$ and $\hat U_m$, respectively. And the average number of users of the tagged $\mu$Wave BS and other $\mu$Wave BS except the tagged $\mu$Wave BS are expressed as $U_\mu$ and $\hat U_\mu$, respectively.
\vspace{-1.0em}
\subsection{Distribution of the distance between the typical user and the serving BS}\vspace{-0.5em}
Unlike the distance between any points in a PPP, which is given as the nearest neighbour distance distribution, this subsection derives the distribution of the distance between the serving BS and the typical user as the conditional probability.

Assume that the distances between the typical user and the serving cache hit and cache miss mmWave SBS with and without the requested $i$th file are denoted by $R_{x_{m_i}^\LOS}$, $R_{x_{m_i}^\NLOS}$, $R_{y_{m_i}^\LOS}$, and $R_{y_{m_i}^\NLOS}$, respectively. Similarly, the distances between the typical user and the serving cache hit and cache miss $\mu$Wave MBS are denoted by $R_{x_{\mu_i}}$ and $R_{y_{\mu_i}}$, respectively. The following Lemma provides the probability density function (PDF) for each of these distances.
\vspace{-0.75em}
\begin{lem1}\label{Distribution of distance between serving BS and typical user}
The PDF of $R_{x_{m_i}^\LOS}$, $R_{x_{m_i}^\NLOS}$, $R_{y_{m_i}^\LOS}$, and $R_{y_{m_i}^\NLOS}$ of SCN and $R_{x_{\mu_i}}$ and $R_{y_{\mu_i}}$ of MCN are given as follows:\vspace{-0.5em}
\begin{align}
f_{R_{\mu_i}} (D)=&\frac{1}{p_{x_{\mu_i}}}  \exp[-\pi \lambda_\mu D^2] \exp[-2\pi \lambda_m Z((\frac{B_m D^{\alpha_\mu}}{B_\mu})^{\frac{1}{\alpha_\LOS}})]\exp[- 2 \pi \lambda_m \hat{Z}((\frac{B_m D^{\alpha_\mu}}{B_\mu})^{\frac{1}{\alpha_\NLOS}}))] 2 \pi \lambda_\mu D p_{\mu_i}\,,\label{distance cache hit}\\
f_{R_{y_{\mu_i}}}(D)=& \frac{1}{p_{y_{\mu_i}}}\exp[-\pi \lambda_\mu D^2] \exp[-2\pi \lambda_m Z((\frac{B_m D^{\alpha_\mu}}{B_\mu})^{\frac{1}{\alpha_\LOS}}))]\exp[- 2 \pi \lambda_m \hat{Z}((\frac{B_m D^{\alpha_\mu}}{B_\mu})^{\frac{1}{\alpha_\NLOS}}))]2\pi\lambda_\mu(1-p_{\mu_i})D\,,\label{distance cache miss}\\
f_{R_{x_{m_i}^\LOS}}(D) =& \frac{1}{p_{x_{m_i}^\LOS}}\exp[-\pi \lambda_\mu (\frac{B_\mu D^{\alpha_\LOS}}{B_m})^{\frac{2}{\alpha_\mu}}] \exp[-2\pi \lambda_m Z(D)]\exp[- 2 \pi \lambda_m \hat{Z}(D^{\frac{\alpha_\LOS}{\alpha_\NLOS}})]2\pi\lambda_m p_{m_i}D e^{-\beta D}\,,\label{distance LOS cache hit}\\
f_{R_{x_{m_i}^\NLOS}}(D) =& \frac{1}{p_{x_{m_i}^\NLOS}}\exp[-\pi \lambda_\mu (\frac{B_\mu D^{\alpha_\NLOS}}{B_m})^{\frac{2}{\alpha_\mu}}] \exp[-2\pi \lambda_m Z(D^{\frac{\alpha_\NLOS}{\alpha_\LOS}})]\exp[- 2 \pi \lambda_m \hat{Z}(D)]2\pi\lambda_m p_{m_i}(D-D e^{-\beta D})\,,\\
f_{R_{y_{m_i}^\LOS}}(D) =& \frac{1}{p_{y_{m_i}^\LOS}}\exp[-\pi \lambda_\mu (\frac{B_\mu D^{\alpha_\LOS}}{B_m})^{\frac{2}{\alpha_\mu}}] \exp[-2\pi \lambda_m Z(D)]\nonumber\\
&\times\exp[- 2 \pi \lambda_m \hat{Z}(D^{\frac{\alpha_\LOS}{\alpha_\NLOS}})]2\pi\lambda_m (1-p_{m_i})D e^{-\beta D}\,,
\\
f_{R_{y_{m_i}^\NLOS}}(D) =&\frac{1}{p_{y_{m_i}^\NLOS}}\exp[-\pi \lambda_\mu (\frac{B_\mu D^{\alpha_\NLOS}}{B_m})^{\frac{2}{\alpha_\mu}}] \exp[-2\pi \lambda_m Z(D^{\frac{\alpha_\NLOS}{\alpha_\LOS}})]\nonumber\\
&\times\exp[- 2 \pi \lambda_m \hat{Z}(D)]2\pi\lambda_m (1-p_{m_i})(D-D e^{-\beta D})\,,\label{distance NLOS cache miss}
\end{align}
where $Z(D)$ and $\hat Z(D)$ are defined in the above and all other parameters are provided before as well.
\end{lem1}
\begin{proof}
The proof of the Lemma can be found in the Appendix \ref{Appears_B}.
\end{proof}
\vspace{-1.0em}
\section{Propagation model}
This section we model the mmWave hybrid beamforming and massive MIMO channels. Particularly, the objective of this section is to illustrate the propagation model. For simplicity, we only consider cache hit association event as the example where the typical user is served by the BS that caches the requested file. Here we assume that the typical user, located at origin, requesting the $i$th file, denoted by $0$, would be associated with either a mmWave SBS located at $x_{m_i}^j$ with $j \in \{\LOS, \NLOS\}$ depending on the LOS or NLOS state or a $\mu$Wave MBS located at $x_{\mu_i}$.
\vspace{-1.0em}
\subsection{MmWave propagation model}\vspace{-0.5em}
The propagation between the typical user and its associated mmWave SBS is via a fully connected hybrid precoder that consists of both RF and BB precoders. However, for simplicity, we assume that each mmWave SBS transmits a total of $U_{x_{m_i}^j}$ streams of data  to serve multiple users but one single stream per user. Therefore, it is sufficient that each user adopts a RF-only combiner with analog beamforming to decode the transmitted single  \cite{7434598}.

\textbf{1) Channel Model}:
The mmWave channel between the serving BS and the typical user, denoted by $\boldsymbol{\mathbf{H}}_{x_{m_i}^j,0}$, is written as\footnote{Hereinafter, for notational simplicity, the typical user subscript $0$ is ignored.}
\begin{align}
\mathbf{H}_{x_{m_i}^j} = \sqrt{\frac{n^m_r n^m_t}{r_{x_{m_i}^j}^{\alpha_j} \eta_{j}}} \sum_{k=1}^{\eta_j} {\mathcal{X}_{k,x_{m_i}^j}} \boldsymbol{\alpha}_{\text{UE}} (\phi_{k,{x_{m_i}^j}})   \boldsymbol{\alpha}_{\text{BS}}^H(\theta_{k,x_{m_i}^j})\,,
\end{align}
where $\mathcal{X}_{k, x_{m_i}^j}$ is the complex channel gain on the $k$th path, distributed as a small-scale Rayleigh fading distribution for both LOS and NLOS paths for analytical tractability \cite{7434598}, $\eta_j$ is the number for paths from the serving BS to the typical user\footnote{As mentioned in \cite{7434598}, this work considers multiple LOS and NLOS paths since {a more general channel model would incorporate scenarios with one or more LOS paths as well as NLOS paths and assume each scatterer provides a single dominant path.}. It is expected that $\eta_\LOS < \eta_\NLOS$ as in \cite{7434598}.}, $\phi_{k,x_{m_i}^j}$ is the angle of arrival (AOA), $\theta_{k, x_{m_i}^j}$ is the angle of departure (AOD), $\alpha_j$ is the the path loss exponent that is different for LOS and NLOS paths. $\boldsymbol{\alpha}_{\text{UE}}(\cdot)$ and $\boldsymbol{\alpha}_{\text{BS}}(\cdot)$ are the array response vectors of each user and BS, respectively, which are modelled as uniform linear arrays (ULAs)\vspace{-1.0em} 
\begin{align}
\boldsymbol{\alpha}_{\text{BS}}(\theta) &= \frac{1}{\sqrt{n^m_t}} [ 1, e^{j \frac{2\pi}{\lambda}d sin(\theta)}, \cdots, e^{(n^m_t-1)j\frac{2\pi}{\lambda}d        sin(\theta)}]\,,\\
\boldsymbol{\alpha}_{\text{UE}}(\phi) &= \frac{1}{\sqrt{n^m_r}}[1, e^{j\frac{2\pi}{\lambda} d{{sin}(\phi)}},\cdots,e^{(n^m_r-1)\frac{2\pi}{\lambda}jdsin(\phi)}]\,,
\end{align}
where $d$ is the distance between antenna elements, commonly is the half of the wavelength ($\lambda$) of the transmitted signal. 

\textbf{2) Received Signal}:
After passing through BB and RF precoders, channel, and RF combiner, the received signal sent from the mmWave BS located at $x_{m_i}^j$ to the typical user is given by\vspace{-0.75em}
\begin{align}\label{received signal mm}
y_0 = & \sqrt{\frac{\mathrm{P}_m}{U_{x_{m_i}^j}}} \bar{\mathbf{h}}_{x_{m_i}^j} \mathbf{v}^{\rm BB}_{x_{m_i}^j} q_{x_{m_i}^j} + \underbrace{\sum\nolimits_{\substack{g \in \mathcal{U}_{x_{m_i}^j}, g \neq 0}}\sqrt{\frac{\mathrm{P}_m}{U_{x_{m_i}^j}}} \bar{\mathbf{h}}_{x_{m_i}^j} \mathbf{v}_{x_{m_i}^j,g}^{\rm BB} q_{x_{m_i}^j,g}}_{\rm IUI} \nonumber \\
& + \underbrace{\sum\nolimits_{b \in \Phi_{m}\backslash\{x_{m_i}^j\}}\sum\nolimits_{t \in \mathcal{U}_{b}}\sqrt{\frac{\mathrm{P}_m}{U_b}} \bar{\mathbf{h}}_{b} \mathbf{v}_{b,t}^{\rm BB} q_{b,t}}_{\rm ICI} + z_0\,,
\end{align}
where the effective channel gain $\boldsymbol{\bar h}_{x_{m_i}^j} = (\mathbf{w}_{x_{m_i}^j}^{\rm RF})^{H} \mathbf{H}_{x_{m_i}^j} \mathbf{V}_{x_{m_i}^j}^{\rm RF}$. The BB and RF precoders are defined as $\mathbf{V}_{x_{m_i}^j}^{\rm BB} = [\mathbf{v}_{x_{m_i}^j,0}^{\rm BB}, \mathbf{v}_{x_{m_i}^j,1}^{\rm BB}, \cdots, \mathbf{v}_{x_{m_i}^j,U_{x_{m_i}^j} - 1}^{\rm BB}]$ and $\mathbf{V}_{x_{m_i}^j}^{\rm RF} = [\mathbf{v}_{x_{m_i}^j,0}^{\rm RF}, \mathbf{v}_{x_{m_i}^j,1}^{\rm RF}, \cdots, \mathbf{v}_{x_{m_i}^j,U_{x_{m_i}^j} - 1}^{\rm RF}]$, respectively. The RF combiner is defined as $ \mathbf{W}_u^{\rm RF} = [\mathbf{w}_{x_{m_i}^j,0}^{\rm RF}, \mathbf{w}_{x_{m_i}^j,1}^{\rm RF}, \cdots, \mathbf{w}_{x_{m_i}^j,U_{x_{m_i}^j}-1}^{\rm RF}]$ where each entity $\mathbf{w}_{x_{m_i}^j,g}^{\rm RF}=[{w}_{g,1}, {w}_{g,2}, \cdots, {w}_{g, n_r^m}]$ with $g \in [0, U_{x_{m_i}^j}-1]$. The transmitted data stream from the mmWave BS $x_{m_i}^j$ is defined as  $Q_{x_{m_i}^j} = [q_{x_{m_i}^j,0}, q_{x_{m_i}^j,1}, \cdots, q_{x_{m_i}^j,U_{x_{m_i}^j}-1}]$. The noise term $z_0 \sim N(0,\sigma_m^2)$. $\frac{\mathrm{P}_m}{U_{x_{m_i}^j}}$ is the average received signal power where the total power $\mathrm{P}_m$ enforces $||\mathbf{V}^{\rm RF}_{x_{m_i}^j} \mathbf{V}^{\rm BB}_{x_{m_i}^j}||^2_F = U_{x_{m_i}^j}$. 

\textbf{3) Design of hybrid precoding}:
Even though we have provided the received signal at the typical user requesting the $i$th file served by the mmWave BS $x_{m_i}^j$, the BB and RF beamforming vectors have not been defined. In order to eliminate the inter user interference  shown in \eqref{received signal mm}, we utilize zero-forcing (ZF) beamforming at the BB precoder, such that $\mathbf{v}_{x_{m_i}^j}^{\rm BB} = (\mathbf{\bar{h}}_{x_{m_i}^j})^{H} \Big(\mathbf{\bar{h}}_{x_{m_i}^j} (\mathbf{\bar{h}}_{x_{m_i}^j})^{H}\Big)^{-1}$.

As for RF precoder and combiner vectors, this work follows a near optimal method in \cite{6292865} to achieve the near-optimal received signal power. As a result, the RF precoding and combing vectors are given by $\mathbf{v}_{x,0}^{\rm RF} = \mathbf{\alpha}_{\rm BS}(\theta_{k_m, x})$ and $\mathbf{w}_{x,0}^{\rm RF} = \mathbf{\alpha}_{\rm UE}(\phi_{k_m, x})$, respectively, where $k_m$ is the path that has the best channel gain \emph{i.e.,}  $k_m = {\rm arg}\max\limits_{k}(\mathcal{X}_{k,x_{m_i}^j})$.   

\textbf{4) SINR characterization}:
Based on \eqref{received signal mm}, the SINR of the typical user from the mmWave BS $x_{m_i}^j$ is formulated as\vspace{-1.0em}
\begin{align}
\text{SINR}_{x_{m_i}^j}^{m} \!\!= \!\!\frac{\frac{\mathrm{P}_m}{U_{x_{m_i}^j}} ||\boldsymbol{\bar h}_{x_{m_i}^j}\, \boldsymbol{v}_{x_{m_i}^j}^{\rm BB}||^2}{\sum\nolimits_{\substack{g\in \mathcal{U}_{x_{m_i}^j}, g\ne 0}} \frac{\mathrm{P}_m}{U_{x_{m_i}^j}} ||\boldsymbol{\bar h}_{x_{m_i}^j,g}\, \boldsymbol{v}_{x_{m_i}^j,g}^{\rm BB} ||^2 + \sum_{b\in \Phi_{m}\backslash\{x_{m_i}^j\}}\sum_{t\in \mathcal{U}_b} \frac{\mathrm{P}_m}{U_b}||\boldsymbol{\bar h}_{b,0} \boldsymbol{v}_{b,t}^{\rm BB}||^2 \!+\! \sigma_m^2} \,. \label{sinr mm}
\end{align} 
However, the analysis on \eqref{sinr mm} is not tractable. Hereinafter, we give a tractable SINR approximation according to \cite{7434598,7421140} with the assumption $n^m_t >> U_{x_{m_i}^j} = U_m$. In particular, the first term in the denominator is zero due to ZF BB precoding and it is reduced to\vspace{-1.50em}
\begin{align}
\text{SINR}_{x_{m_i}^j}^{m} \approx \frac{\frac{\mathrm{P}_m}{U_{x_{m_i}^j}} \frac{n_t^m n_r^m }{\eta_{j}} \mathcal{X}_{x_{m_i}^j}^2 r_{x_{m_i}^j}^{-\alpha_{j}} p_{\rm ZF}}{I_{x_{m_i}^j} + \sigma_m^2}\,,\label{sinr mm approximation}
\end{align}
where using the ON/OFF approximation model for the array response vectors\footnote{Assume $n^m_t >> U_{x_{m_i}^j}$, the array response model enhances the analysis tractability since $\mathbf{\alpha}^H(\theta_1) * \mathbf{\alpha}(\theta_2) = 0$ if $\theta_1 \neq \theta_2$. Otherwise, it is a nonzero value but lower bounded by $0$ due to ZF precoding.}, $p_{\rm ZF}$ is the ZF precoding penalty that is the probability that the signal power is mainly dominant at the typical user and the signal powers of other users in the tagged cell are lower bounded by $0$,  given by \cite{7421140}\vspace{-2.5em}
\begin{align}
p_{ZF} = \left\{\begin{array}{ll} 1, & {\rm w.p. } \,\,(1 - \frac{1}{n^m_r})^{{U}_{x_{m_i}^j}-1}\\0, & {\rm otherwise }\end{array}\right..
\end{align}
The second term in the denominator, $I_{x_{m_i}^j}$, the inter-cell-interference, given by\vspace{-1.0em}
\begin{align} \label{ICI mm}
I_{x_{m_i}^j} = &\sum\nolimits_{\hat{j}\in\{\LOS, \NLOS\}}\sum\nolimits_{\substack{b \in \Phi_m^{\hat{j}}\,,\; b\neq x_{m_i}^j}} \frac{\mathrm{P}_m}{{U}_{b}}\frac{n^m_rn^m_t}{\eta_{\hat j}}r_{b}^{-\alpha_{\hat{j}}} \\
&\times\sum\nolimits_{t\in \mathcal{U}_{b}} \parallel\sum\nolimits_{k=1}^{\eta_{\hat{j}}}{\mathcal{X}_{k,b}}\underbrace{ \mathbf{\alpha}_{\text{UE}}^H(\phi_{x_{m_i}^j}) \mathbf{\alpha}_{\text{UE}}(\phi_{k,b})  \mathbf{\alpha}_{\text{BS}}^H(\theta_{k,b})  \mathbf{\alpha}_{\text{BS}}(\theta_{b,t})}_{\gamma_{k,b,t}}\parallel^2\,.\nonumber
\end{align}
However, unlike the fact that IUI in the tagged cell is cancelled by ZF precoding, there is no ZF penalty on any of the interfering signals from other BSs except the serving BS. Therefore, ON/OFF model approximation made for the  IUI is not suitable and accurate to ICI. Instead of setting $0$ to the unaligned interfering signal power, we give another values $1> \rho_{\rm UE} > 0$ and $1 >\rho_{\rm BS} > 0 $ to approximate the term $\gamma_{k,b,t}$ by\vspace{-0.75em}
\begin{align}
\gamma_{k,b,t} = \left\{\begin{array}{ll}1,&{ \rm if}\,\,\,\phi_{x_{m_i}^j}=\phi_{k,b},\theta_{k,b}=\theta_{b,t}\\\rho_{\rm BS},&{\rm if}\,\,\,\phi_{x_{m_i}^j}\neq\phi_{k,b},\theta_{k,b}=\theta_{b,t}\\\rho_{\rm UE},&{\rm if}\,\,\,\phi_{x_{m_i}^j}=\phi_{k,b},\theta_{k,b}\neq\theta_{b,t}\\\rho_{\rm BS}\rho_{\rm UE},& {\rm otherwise }\end{array}\right.\label{inter}\,.
\end{align}
Further, \eqref{ICI mm} is reduced to a simple closed-form expression that will be used in the Performance Metric section.
\vspace{-1.65em}
\subsection{Massive MIMO}\vspace{-0.5em}
\textbf{Received signal}:
For massive MIMO access link, this work applies ZF beamforming with equal power allocation to the massive MIMO enabled MBSs. Besides, the massive MIMO enabled MBSs performs the average channel estimation within TDD mode to acquire the channel state information (CSI) while not performing any channel estimation at users due to the channel reciprocity in TDD mode, which users do not have any channel knowledge. Unlike using instantaneous CSI estimation to calculate the normal SINR measure, this work considers the worst Gaussian noise as a lower bounding technique for any type of precoding strategies \cite{5898372}. Consequently, the additional self-interference caused by the average channel estimation appears in the SINR denominator and considered as part of the noise. However, the method in \cite{5898372} did not utilize stochastic geometry while this work considers the dense cellular network from the stochastic geometric framework. Based on the instantaneous received signal expression rewritten to compute the achievable data rate \cite{5898372, 8424570}, the received signal at the typical user requesting the $i$th file is shown as\vspace{-0.5em}
\begin{align}\label{recieved signal mu}
y_{x_{\mu_i}} = &\sqrt{\frac{\mathrm{P}_\mu}{U_{x_{\mu_i}}}} \mathbb{E}[{\hat{H}}_{x_{\mu_i}}] q_{x_{\mu_i}} r_{x_{\mu_i}}^{-\frac{\alpha_\mu}{2}} + \sqrt{\frac{\mathrm{P}_\mu}{U_{x_{\mu_i}}}}({\hat{H}}_{x_{\mu_i}} - \mathbb{E}[{\hat{H}}_{x_{\mu_i}}]) q_{x_{\mu_i}} r_{x_{\mu_i}}^{-\frac{\alpha_\mu}{2}}  \\
&+\underbrace{\sum\nolimits_{\substack{g \in \mathcal{U}_{x_{\mu_i}},\;g \neq 0}}\sqrt{ \frac{\mathrm{P}_\mu}{U_{x_{\mu_i}}}} {\hat{H}}_{x_{\mu_i}}  q_{x_{\mu_i},g} r_{x_{\mu_i},y}^{-\frac{\alpha_\mu}{2}}}_{IUI}
+\underbrace{\sum\nolimits_{b \in \Phi_{\mu}\backslash \{x_{\mu_i}\}} \sum\nolimits_{t\in\mathcal{U}_b} \sqrt{\frac{\mathrm{P}_\mu}{U_b}} {\hat{H}}_{b,t}  q_{b,t} r_{b,t}^{-\frac{\alpha_\mu}{2}}}_{ICI} + z_0\,, \nonumber
\end{align}
where $q_{x_{\mu_i}}$ is the transmitted signal from the MBS at $x_{\mu_i}$ to the user at the origin, $|\hat{H}_{x_{\mu_i}}|^2 = \mathcal{G}_{x_{\mu_i}} \thicksim \Gamma(n_t^\mu - U_{x_{\mu_i}} +1, 1)$ is the channel gain among the serving $\mu$Wave MBS located at $x_{\mu_i}$ and the user located at the origin. $(\sum_{t \in \mathcal{U}_b} \hat{H}_{b,t} )^2 = \mathcal{G}_{b} \thicksim \Gamma(U_b, 1)$ is the channel gain between the MBS except the serving MBS to the typical user. $r_{x_{\mu_i}}$ is the distance between the serving MBS and the typical user. $z_0 \thicksim N(0,\sigma_\mu^2)$ is the noise term. 

\textbf{SINR characterization}:
Based on \eqref{recieved signal mu}, the SINR is given by
\begin{align}
\text{SINR}^{\rm ZF}_{x_{\mu_i}} = \frac{\frac{\mathrm{P}_\mu}{U_{x_{\mu_i}}} (\mathbb{E}[\sqrt{\mathcal{G}_{x_{\mu_i}}}])^2 r_{x_{\mu_i}}^{-\alpha_\mu}}{\frac{\mathrm{P}_\mu}{{U}_{x_{\mu_i}}}(\sqrt{\mathcal{G}_{x_{\mu_i}}}-\mathbb{E}[\sqrt{\mathcal{G}_{x_{\mu_i}}}] )^2 r_{x_{\mu_i}}^{-\alpha_\mu} + \sum_{\substack{b\in\Phi_{\mu}\\b\ne x_{\mu_i}}} \frac{\mathrm{P}_\mu}{{U}_{b}} \mathcal{G}_{b} r_{b}^{-\alpha_\mu} + \sigma_\mu^2}\,,\label{SINR mu}
\end{align}
In the next performance metric section, we will use \eqref{SINR mu} to compute the ASP of file delivery. In particular, this work considers 
\vspace{-1.0em}
\section{Performance Metric}\vspace{-0.75em}
The QoS is closely related to the ASP of file delivery, the delay experienced by users, network capacity, and the backhaul load. However, it is evident that the downlink transmission over the wireless medium incurs outage and delay mainly due to the interference from concurrent transmissions and channel fading. Therefore, we consider a simple retransmission protocol    where a packet of requested content is repeatedly transmitted until its successful delivery, up to a pre-defined number of retransmission attempts $N$. Indeed, inferring whether a packet delivery is successful or not at the BS essentially relies on the SINR being higher than the predefined threshold $\nu_i$. If a packet is delivered successfully, we shall assume that the BS receives a one-bit acknowledgement message from the UE with negligible delay and error. Otherwise, if the delivery fails, the BS receives a one-bit negative acknowledge message in the same vein. An outage event occurs if the packet is not delivered after $N$ attempts. Therefore, this work considers three retransmission based performance metrics -- retransmission based ASP of the file delivery, average packet delay, throughput per user, and the backhaul load per unit area. In particular, we consider the static scenario where the locations of users and BSs are stationary in the $N$ consecutive retransmission attempts. The channel fading power coefficient is considered stationary and i.i.d. in each attempt slot. For analytical tractability and simplicity, we neglect the temporal interference correlation and every attempt is an independent event to give the best-case retransmission based ASP of file delivery (upper bound), the packet delay (lower bound), throughput per user(upper bound), and the backhaul load per area (upper bound). For more details about the complete characterization of retransmission based ASP of file delivery and latency, the readers are recommended to look into \cite{7536893} and \cite{8377141}.  
\vspace{-1.25em}
\subsection{Retransmission based ASP of file delivery}\vspace{-0.25em}
Due to the independence of PPPs $\Phi_m$ and $\Phi_\mu$, the total 6 association events are considered as independent and in each event the retransmission protocol is implemented. Therefore, according to the law of total probability and the conditional probability, the retransmission based the ASP of file delivery is given by\vspace{-1.0em}
\begin{align}\label{retransmission ASP}
\mathcal{P}^{\rm (re)}_s(\nu_i, N) =& \sum\nolimits_{i=1}^{F} f_i (p_{x_{m_i}^\LOS} \mathcal{P}^{\rm (re)}_{x_{m_i}^\LOS} + p_{x_{m_i}^\NLOS} \mathcal{P}^{\rm (re)}_{x_{m_i}^\NLOS} + p_{y_{m_i}^\LOS} \mathcal{P}^{\rm (re)}_{y_{m_i}^\LOS} + p_{y_{m_i}^\NLOS} \mathcal{P}^{\rm (re)}_{y_{m_i}^\NLOS} + p_{x_{\mu_i}} \mathcal{P}^{\rm (re)}_{x_{\mu_i}} + p_{y_{\mu_i}} \mathcal{P}^{\rm (re)}_{y_{\mu_i}})
\end{align}
As for the cache hit, the ASP of file delivery is only associated with access link. As for the cache miss, the ASP of file delivery is connected with both access and backhaul ASP of file delivery. Now considering the probability that the serving BS storing the requested content successfully transmits the packet of the requested content in a single attempt referred to as $p_s$, the probability that, in at least one of the $N$ time slots, the UE is scheduled and the transmission succeeds is given as $1 - (1 - p_s)^N$. Therefore, it is necessary to first compute the ASP of file delivery in a single attempt in each association event.

$1)$ \textit{The ASP of file delivery of the mmWave SBS in a single attempt}: The ASP of file delivery in a single attempt is defined as the probability that each file requested by the typical user must be transmitted by the serving BS located at $x$ supporting the data rate of $(W/U) \text{log}(1 + \rm SINR)$ over the target data rate $\nu_i$ (bits/s), that is formulated by\vspace{-1.0em}
\begin{equation}
\mathcal{P}_{x}(\nu_i) = \mathbb{P}[\mathcal{R}_{x} > \nu_i]\,.\label{ASP of file delivery}\vspace{-0.5em}
\end{equation}
Assume the bandwidths of SCN and MCN are $W_m$ and $W_\mu$, respectively, the data rate of the tagged mmWave located at $x_m$ and $\mu$Wave located at $x_\mu$ are given as \vspace{-1.0em}
\begin{align}
\mathcal{R}_{x_m} &= \frac{W_m}{U_{x_m}} \text{log}(1 + \rm SINR_{x_m})\,, \\
\mathcal{R}_{x_\mu} &= \frac{W_\mu}{U_{x_\mu}} \text{log}(1 + \rm SINR_{x_\mu})\,.
\end{align}

Based on the definition of ASP of file delivery, the following proposition \ref{ASP of file delivery mm} gives the ASP of file delivery of the associated mmWave SBS in a single attempt given the distance between the serving BS and the typical user.
\begin{prop1}
The conditional ASP of file delivery of mmWave SBS storing the requested $i$th file located at $x_{m_i}^\LOS$ and $x_{m_i}^\NLOS$ in LOS and NLOS transmissions are lower bounded as\vspace{-0.5em}
\begin{align}\label{ASP of file delivery mm}
\mathcal{P}_{s}^{({\rm hit}, \LOS)}(\nu_i)&\geq(1 - \frac{1}{n^m_r})^{({U}_{m} - 1)}  \exp\Big(\frac{-Q_i \sigma^2_m }{{G}_{x_{m_i}^{\LOS}} R_{x_{m_i}^{\LOS}}^{-\alpha_{\LOS}}}\Big)\nonumber \\
&\times\exp\Big[\int_{0}^{R_{x_{m_i}^{\LOS}}} \Big[1 - \Big({1 + s \mathrm{P}_m n^m_t n^m_r r^{-\alpha_\LOS} }\Big)^{-\eta_{\LOS}}\Big]2 \pi \lambda_m p_\LOS(r) r p_{m_i} \textup{d} r\Big]\nonumber \\
&\times\exp\Big[-\sum\nolimits_{j\in\{\LOS,\NLOS\}}\int_{0}^{\infty} \Big[1 - \Big({1 +s \mathrm{P}_m n^m_t n^m_r r^{-\alpha_j} }\Big)^{-\eta_{j}}\Big]2 \pi \lambda_m p_j(r) r  \textup{d} r\Big]\,,
\end{align}
where $G_{x_{m_i}^{\LOS}} = \frac{\mathrm{P}_m}{{U}_{m}}\frac{n^m_r n^m_t}{\eta_{{\LOS}}}$ and $s = \frac{-Q_i}{{G}_{x_{m_i}^{\LOS}}R_{x_{m_i}^{\LOS}}^{-\alpha_\LOS}}$. $Q_i = 2 ^{\frac{\nu_i U_m}{W_m} } - 1$.\vspace{-1.0em}
\begin{align}
\mathcal{P}_{s}^{({\rm hit}, \NLOS)}(\nu_i)&\geq(1 - \frac{1}{n^m_r})^{({U}_{m} - 1)}  \exp\Big(\frac{-Q_i \sigma^2_m }{{G}_{x_{m_i}^{\NLOS}} R_{x_{m_i}^{\NLOS}}^{-\alpha_{\NLOS}}}\Big)\nonumber \\
&\times \exp\Big[\int_{0}^{R_{x_{m_i}^{\NLOS}}} \Big[1 - \Big({1 - {s \mathrm{P}_m n^m_r n^m_t r^{-\alpha_\NLOS} }}\Big)^{-\eta_{\NLOS}}\Big]2 \pi \lambda_m p_\NLOS(r) r p_{m_i} \textup{d} r\Big]\nonumber \\
&\times\exp\Big[-\sum_{j\in\{\LOS,\NLOS\}}\int_{0}^{\infty} \Big[1 - \Big({1 - {s \mathrm{P}_m n^m_r n^m_t r^{-\alpha_j} }}\Big)^{-\eta_{j}}\Big]2 \pi \lambda_m p_j(r) r  \textup{d} r\Big]\,,
\end{align}
where $G_{x_{m_i}^{\NLOS}} = \frac{\mathrm{P}_m}{{U}_{m}}\frac{n^m_r n^m_t}{\eta_{{\NLOS}}}$ and $s = \frac{-Q_i}{{G}_{x_{m_i}^{\NLOS}}R_{x_{m_i}^{\NLOS}}^{-\alpha_\NLOS}}$. The conditional ASP of file delivery of mmWave SBS  not storing the requested $i$th file located at $y_{m_i}^\LOS$ and $y_{m_i}^\NLOS$ in LOS and NLOS transmissions are lower bounded as\vspace{-1.0em}
\begin{align}
\mathcal{P}_{s}^{({\rm miss}, \LOS)}(\nu_i)&\geq(1 - \frac{1}{n^m_r})^{({U}_{m} - 1)} \exp\Big(\frac{-Q_i \sigma^2_m }{{G}_{y_{m_i}^{\LOS}} R_{y_{m_i}^{\LOS}}^{-\alpha_{\LOS}}}\Big)\nonumber \\
&\times \exp\Big[-\int_{0}^{R_{y_{m_i}^{\LOS}}} \Big[1 - \Big({1 - {s \mathrm{P}_m n^m_r n^m_t r^{-\alpha_\LOS} }}\Big)^{-\eta_{\LOS}}\Big]2 \pi \lambda_m p_\LOS(r) r p_{m_i} \textup{d} r\Big]\nonumber \\
&\times\exp\Big[-\int_{0}^{\infty} \Big[1 - \Big({1 - {s \mathrm{P}_m n^m_r n^m_t r^{-\alpha_\NLOS} }}\Big)^{-\eta_{\NLOS}}\Big]2 \pi \lambda_m p_\NLOS(r) r  \textup{d} r\Big]\nonumber \\
&\times\exp\Big[-\int_{R_{y_{m_i}^{\LOS}}}^{\infty} \Big[1 - \Big({1 - {s \mathrm{P}_m n^m_r n^m_t r^{-\alpha_\LOS} }}\Big)^{-\eta_{\LOS}}\Big]2 \pi \lambda_m p_\LOS(r) r  \textup{d} r\Big]\,,
\end{align}
where $G_{y_{m_i}^{\LOS}} = \frac{\mathrm{P}_m}{{U}_{m}}\frac{n^m_r n^m_t}{\eta_{{\LOS}}}$ and $s = \frac{-Q_i}{{G}_{y_{m_i}^{\LOS}}R_{y_{m_i}^{\LOS}}^{-\alpha_\LOS}}$.\vspace{-1.0em}
\begin{align}
\mathcal{P}_{s}^{({\rm miss}, \NLOS)}(\nu_i)&\geq(1 - \frac{1}{n^m_r})^{({U}_{m} - 1)}  \exp\Big(\frac{-Q_i \sigma^2_m }{{G}_{y_{m_i}^{\NLOS}} R_{y_{m_i}^{\NLOS}}^{-\alpha_{\NLOS}}}\Big)\nonumber \\
&\times\exp\Big[-\int_{0}^{R_{y_{m_i}^{\NLOS}}} \Big[1 - \Big({1 - {s \mathrm{P}_m n^m_r n^m_t r^{-\alpha_\NLOS} }}\Big)^{-\eta_{\NLOS}}\Big]2 \pi \lambda_m p_\LOS(r) r p_{m_i} \textup{d} r\Big]\nonumber \\
&\times\exp\Big[-\int_{0}^{\infty} \Big[1 - \Big({1 - {s \mathrm{P}_m n^m_r n^m_t r^{-\alpha_\LOS} }}\Big)^{-\eta_{\LOS}}\Big]2 \pi \lambda_m p_\NLOS(r) r  \textup{d} r\Big]\nonumber \\
&\times\exp\Big[-\int_{R_{y_{m_i}^{\NLOS}}}^{\infty} \Big[1 - \Big({1 - {s \mathrm{P}_m n^m_r n^m_t r^{-\alpha_\NLOS} }}\Big)^{-\eta_{\NLOS}}\Big]2 \pi \lambda_m p_\NLOS(r) r  \textup{d} r\Big]\,,
\end{align}
where $G_{y_{m_i}^{\NLOS}} = \frac{\mathrm{P}_m}{{U}_{m}}\frac{n^m_r n^m_t}{\eta_{y_{m_i}^{\NLOS}}}$ and $s = \frac{-Q_i}{{G}_{y_{m_i}^{\NLOS}}R_{y_{m_i}^{\NLOS}}^{-\alpha_\NLOS}}$.
\end{prop1}
\begin{proof}
The proof of the Theorem is given in the Appendix \ref{Appears_C}.
\end{proof}
The above gives the lower bound of conditional ASP of file delivery in SCN and the upper bound is acquired through the change of $\eta_j = 1$ with $j \in \{\LOS, \NLOS\}$ and put term $(\rho_{\rm UE} \rho_{\rm BS})^2$ in the above equations as shown in Appendix \ref{Appears_A}.

$2)$ \textit{ASP of file delivery of the $\mu$Wave MBS in a single attempt}: Before computing the conditional ASP of file delivery of cache hit and cache miss $\mu$Wave MBSs, we initially provide the achievable data rate and then use it to compute the conditional ASP of file delivery.\vspace{-1.25em}
\begin{lem1}\label{achievable data rate mu}
The achievable data rate of cache hit and cache miss $\mu$Wave BSs are respectively given as\vspace{-1.0em}
\begin{align}
\mathcal{\bar R}_{x_{\mu}} &=\mathbb{E}[\mathcal{R}_{x_{\mu}}] = \frac{W_\mu}{U_{\mu}} \text{log}\Big(1 + \frac{C_1 r_{x_{\mu}}^{-\alpha_\mu}}{C_2 r_{x_{\mu_i}}^{-\alpha_\mu} + C'_3 r_{x_{\mu}}^{-\alpha_\mu+2} +  C''_3+\sigma_\mu^2}\Big)\,,
\end{align} 
where \vspace{-1.50em}
\begin{align}
C_1 &= \frac{\mathrm{P}_\mu}{{U}_{\mu}}\Big(\frac{\Gamma(n_t^\mu - {U}_{\mu} + \frac{3}{2})}{\Gamma(n_t^\mu - {U}_{\mu} + 1)}\Big)^2,\;&C_2 &= \frac{\mathrm{P}_\mu}{{U}_{\mu}} \Big(n_t^\mu - {U}_{\mu} + 1 \Big)- C_1,\\
C'_3 &= {\mathrm{P}_\mu 2 \pi \lambda_\mu p_{\mu_i}  \frac{1}{\alpha_\mu - 2}},\;&C''_3 &= \mathrm{P}_\mu 2 \pi \lambda_\mu (1-p_{\mu_i})  \frac{1}{\alpha_\mu - 2}.
\end{align}
\begin{align}
\mathcal{\bar R}_{y_{\mu}}& =\mathbb{E}[\mathcal{R}_{y_{\mu}}] = \frac{W_\mu}{U_\mu} \text{log}\Big(1 + \frac{C_1 r_{y_{\mu}}^{-\alpha_\mu}}{C_2 r_{y_{\mu_i}}^{-\alpha_\mu} + C'_3 r_{y_{\mu}}^{-\alpha_\mu+2} +  C''_3+\sigma_\mu^2}\Big)\,,
\end{align}
where \vspace{-1.50em}
\begin{align}
C_1 &= \frac{\mathrm{P}_\mu}{{U}_{\mu}}\Big(\frac{\Gamma(n_t^\mu - {U}_{\mu} + \frac{3}{2})}{\Gamma(n_t^\mu - {U}_{\mu} + 1)}\Big)^2,\;&C_2 &= \frac{\mathrm{P}_\mu}{{U}_{\mu}} \Big(n_t^\mu - {U}_{\mu} + 1 \Big)- C_1,\\
C'_3 &= {\mathrm{P}_\mu 2 \pi \lambda_\mu (1-p_{\mu_i})  \frac{1}{\alpha_\mu - 2}},\;&C''_3 &= \mathrm{P}_\mu 2 \pi \lambda_\mu p_{\mu_i}  \frac{1}{\alpha_\mu - 2}.
\end{align}
\end{lem1}
\begin{proof}
The proof is offered in Appendix \ref{Appears_D}.
\end{proof}
Using the Lemma \ref{achievable data rate mu}, we give the conditional ASP of file delivery in MCN.
\begin{prop1}\label{Prop_conditional_ASP_mu}
The conditional ASP of file delivery of $\mu$Wave MBS storing and not storing the requested $i$th file located at $x_{\mu_i}$ and $y_{\mu_i}$ are respectively lower bounded as\vspace{-1.50em}
\begin{align}
\mathcal{P}_{x_{\mu_i}}^{(\rm hit)}(\nu_i)&=\mathbb{P}[\mathcal{\bar R}_{x_{\mu_i}} \geq \nu_i]\geq \mathbb{P}\Big[r_{x_{\mu_i}} \leq \floor{R^*_{x_{\mu_i}}}\Big]\nonumber \\
&=\int_{1}^{\floor{R^*_{x_{\mu_i}}}} f(r_{x_{\mu_i}}) \textup{d} r_{x_{\mu_i}}\,,
\end{align}
and \vspace{-1.75em}
\begin{align}
\mathcal{P}_{y_{\mu_i}}^{(\rm miss)}(\nu_i) &= \mathbb{P}[\mathcal{\bar R}_{y_{\mu_i}} \geq \nu_i]\geq \mathbb{P}\Big[r_{y_{\mu_i}} \leq \floor{R^*_{y_{\mu_i}}}\Big]\nonumber \\
&=\int_{1}^{\floor{R^*_{y_{\mu_i}}}} f(r_{y_{\mu_i}}) \textup{d} r_{y_{\mu_i}}\,,
\end{align}
Their upper bounds are also given through the change of $\floor{R^*}$ for $\ceil{R^*}$ as show in Appendix \ref{Appears_E}.
\end{prop1}
Now we give the ASP of file delivery on backhaul links. Initially, users asynchronously request files from $\mathcal{F}$ and all asynchronous requests for a specific file that is not cached in the local caches will lead to the redundant transmission and load on backhaul links. In order to reflect this phenomenon, we formulate the backhaul ASP of file delivery that is the function of the cell load by assuming that each user is equally allocated the same part of total backhaul capacity. Since the QoS requirement requires the minimum data rate i.e., target data rate,  we assume that all users are equally allocated the backhaul capacity $\nu_i$. Due to the limited backhaul capacity, the maximum number of users served  by the backhaul links is fixed as $N_b = \frac{C_b}{ \nu_i}$ with equal target rate for all files. Denote the number of cache miss users as $N_{\rm miss}$. If $N_{\rm miss} \leq N_b$, all the cache miss users can be scheduled with the backhaul link. On the contrary, if $N_{\rm miss} > N_b$, the backhaul link fails to support all the cache miss users and will randomly pick $N_b$ users with equal probability $\frac{N_b}{N_{\rm miss}}$.  

{Assuming that the requested $i$th file by the typical user is not cached in the associated BS and there are another $n$ cache miss users requesting the $i$th file associated with the tagged BS, the ASP of file delivery on backhaul link of each mmWave BS is given by}\vspace{-1.0em}
\begin{align}\label{full cache backhaul mm}
\mathcal{P}_b^m(\nu_i) &= \sum\nolimits_{n=0}^{N_b - 1} \binom{U_m-1}{n} p_{\rm hit,m}^{N_b - n - 1} {(1- p_{\rm hit,m})}^{n}  \nonumber \\
&+ \sum\nolimits_{n= N_b}^{U_b-1} \binom{U_m-1}{n} (p_{\rm hit})^{U_m-n-1} (1-p_{\rm hit})^{n} \frac{N_b}{(n+1)}\nonumber \\
&=\sum\nolimits_{n=0}^{U_m-1}\binom{U_m-1}{n} p_{\rm hit,m}^{U_m-n-1} (1-p_{\rm hit,m})^{n} \min\Big\{1,\frac{N_b}{(n+1)}\Big\}\,,
\end{align}
where $p_{\rm hit, m} = \sum_{i=1}^{F} f_i p_{m_i}$ is the average probability that the files requested by all other users are cached in the local caches of the tagged mmWave BS, which does not cause the backhaul load. In particular, when the total number of cache miss users $N_{\rm miss}$ is greater than or equal to $N_b$, each user has equal probability $\frac{N_b}{N_{\rm miss}}$.
In a similar way, the backhaul ASP of file delivery on backhaul links of each $\mu$Wave BS is acquired as\vspace{-1.0em} 
\begin{align}\label{full cache backhaul mu}
\mathcal{P}_b^\mu(\nu_i) &= \sum\nolimits_{n=0}^{N_b - 1} \binom{U_\mu-1}{n} (p_{\rm hit,\mu})^{N_b - n - 1} (1-p_{\rm hit,\mu})^{n}  \nonumber \\
&+ \sum\nolimits_{n= N_b}^{U_\mu-1} \binom{U_\mu-1}{n} (p_{\rm hit, \mu})^{U_\mu-n-1} (1-p_{\rm hit,\mu})^{n} \frac{N_b}{(n+1)}\nonumber \\
&=\sum\nolimits_{n=0}^{U_\mu-1}\binom{U_\mu-1}{n}(p_{\rm hit,\mu})^{U_\mu-n-1} (1-p_{\rm hit,\mu})^{n} \min\Big\{1,\frac{N_b}{(n+1)}\Big\}\,,
\end{align}
where $p_{\rm hit,\mu} = \sum_{1}^{F} f_i p_{\mu_i}$ is the average probability that files requested by all other users are cached in the tagged $\mu$Wave BS.

$3)$ \textit{The retransmission based ASP of file delivery}: As for the SCN, the retransmission based ASP of file delivery $\mathcal{P}^{(re)}_{x_{m_i}^\LOS}$, $\mathcal{P}^{(re)}_{x_{m_i}^\NLOS}$, $\mathcal{P}^{(re)}_{y_{m_i}^\LOS}$, $\mathcal{P}^{(re)}_{y_{m_i}^\NLOS}$ are respectively given as\vspace{-0.75em}
\begin{align}
\mathcal{P}^{(re)}_{x_{m_i}^\LOS} &= \int_{0}^{\infty} [1 - (1 - \mathcal{P}_{x_{m_i}^\LOS}^{(\rm hit, \LOS)} (R_{x_{m_i}^\LOS}))^{N}] f(R_{x_{m_i}^\LOS}) \textup{d} R_{x_{m_i}^\LOS} \,,\\
\mathcal{P}^{(re)}_{x_{m_i}^\NLOS} &= \int_{0}^{\infty} [1 - (1 - \mathcal{P}_{x_{m_i}^\NLOS}^{(\rm hit, \NLOS)} (R_{x_{m_i}^\NLOS}))^{N}]f(R_{x_{m_i}^\NLOS}) \textup{d} R_{x_{m_i}^\NLOS} \,,\\
\mathcal{P}^{(re)}_{y_{m_i}^\LOS} &= \int_{0}^{\infty} [1 - (1 - \mathcal{P}_{y_{m_i}^\LOS}^{(\rm miss, \LOS)} (R_{y_{m_i}^\LOS}) \mathcal{P}_b^m)^{N}] f(R_{y_{m_i}^\LOS}) \textup{d} R_{y_{m_i}^\LOS} \,,\\
\mathcal{P}^{(re)}_{y_{m_i}^\NLOS} &= \int_{0}^{\infty} [1 - (1 - \mathcal{P}_{y_{m_i}^\NLOS}^{(\rm miss, \NLOS)} (R_{y_{m_i}^\NLOS}) \mathcal{P}_b^m )^{N}] f(R_{y_{m_i}^\NLOS}) \textup{d} R_{y_{m_i}^\NLOS} \,,
\end{align}
where $f(\cdot)$ is the related distribution of the distance between the serving SBS and the typical user given by \eqref{distance LOS cache hit} -- \eqref{distance NLOS cache miss}. 

However, due to the massive MIMO channel hardening, the achievable data rate based ASP of file delivery is not a function of the distance.  The retransmission based ASP of file delivery $\mathcal{P}^{(\rm re)}_{x_{\mu_i}}$ and $\mathcal{P}^{(\rm re)}_{y_{\mu_i}}$ are given directly as\vspace{-1.50em}
\begin{align}
\mathcal{P}^{(\rm re)}_{x_{\mu_i}} &= 1 - (1 - \mathcal{P}_{x_{\mu_i}}^{\rm (hit)})^N\,,\\
\mathcal{P}_{y_{\mu_i}}^{(\rm re)} &= 1 - ( 1- \mathcal{P}_{y_{\mu_i}}^{(\rm miss)} \mathcal{P}^\mu_b)^N\,.
\end{align}

Finally, by substituting the related functions into \eqref{retransmission ASP}, we have the desirable result.
\vspace{-1.25em}
\subsection{Latency}\vspace{-0.25em}
This section we study the average latency that is composed of delay in access link, backhaul link, and caches. In fact, as analysed in \cite{zhang2016fundamentals}, the packet delay has a significant impact on the queueing and end-to-end delays, and it consists of the packet transmission and propagation delays along the link as well as the processing delay at each node. However, due to less delay in caches compared to the others, we ignore the caching delay that is generated when serving a user by fetching the content from the local cache.

\begin{itemize}
\item Delay in backhaul link: The packet delay mainly comes from the processing time and transmission delay, which means that the propagation delays can be ignored due to the highly reliable transmission of the wired backhaul.

\item Delay in access link: The packet delay along the link mainly comes from the transmission time.
\end{itemize}

Considering the aforementioned sources of delay, the average delay experienced by the typical user can be defined as \vspace{-1.0em}
\begin{align}
D =& \sum\nolimits_{i=1}^{F} f_i (p_{x_{m_i}^\LOS} D_{x_{m_i}^\LOS} + p_{x_{m_i}^\NLOS} D_{x_{m_i}^\NLOS} + p_{y_{m_i}^\LOS} D_{y_{m_i}^\LOS} + p_{y_{m_i}^\NLOS} D_{y_{m_i}^\NLOS} + p_{x_{\mu_i}} D_{x_{\mu_i}} + p_{y_{\mu_i}} D_{y_{\mu_i}})
\end{align}
Now we derive $D_{x_{m_i}^\LOS}$. Consider that the delay by a single transmission is referred to as $T_0$ and the average delay is at least $T_0$ with the probability 1. Given the first failure with probability $1 - \mathcal{P}_{x_{m_i}^\LOS}(R_{x_{m_i}^\LOS})$, the second retransmission generates another $T_0$ additional time. In this manner, the averaged delay $D_{x_{m_i}^\LOS}$ is given by \cite{7041201}\vspace{-1.0em}
\begin{align}
D_{x_{m_i}^\LOS} &= T_0 + (1 - \mathcal{P}_{x_{m_i}^\LOS}(R_{x_{m_i}^\LOS}))[T_0 + (1-\mathcal{P}_{x_{m_i}^\LOS}(R_{x_{m_i}^\LOS}))(T_0 + \cdots)]\nonumber \\
&\overset{(a)}{=}T_0 \int_{0}^{\infty}\frac{1 - (1 - \mathcal{P}_{x_{m_i}^\LOS}(R_{x_{m_i}^\LOS}))^N}{\mathcal{P}_{x_{m_i}^\LOS}(R_{x_{m_i}^\LOS})}f(R_{x_{m_i}^\LOS}) \textup{d} R_{x_{m_i}^\LOS}\,,
\end{align}
where $(a)$ follows from the law of total expectation with respect to distance. Similarly, all other terms can be derived. The main difference of the derivation between the cache hit and cache miss is that $1)$ the successful probability of the single slot and $2)$ the time $T_0$ for one transmission attempt.
As for cache hit event, there is only the access ASP of file delivery in a single attempt while we need to consider backhaul ASP along with access ASP of file delivery together for cache miss. In addition, $T_0$ is only the access packet transmission time for cache hit event but also backhaul transmission and processing time as well for cache miss event. In the following, we respectively give these delays.
\begin{itemize}
\item Latency in access link: Since we only consider transmission delay and the packet delay is proportional to the file size\footnote{Usually it is not possible to send the whole file at a time and each file should be divided into small sized chunks/packets. However, for the case where the node locations remain fixed, we consider the whole file size as the packet size and the channel keeps stationary in the transmission time, namely the Doppler effect should be ignored.}, $T_0 = \frac{S}{\bar{\mathcal{R}}}$ is treated as the time consumption in one transmission.

\item Latency in backhaul link: As for the transmission delay, since each user is allocated the same backhaul capacity (the target rate), it is defined as $\frac{S}{\nu_i}$. As for the processing time, we model the mean packet processing delay distributed as gamma distribution including the processing time generated in gateway and relay hubs as given in \cite{zhang2016fundamentals}, where the processing delay is given by $(\frac{\lambda_{j}}{\lambda_g}{k}_1 + (\frac{1}{r\sqrt{2\lambda_g}} -1){k}_2 )(a + S \omega)$ with $j \in \{m,\mu\}$. $a, \omega, k_1$ and $k_2$ are the constants that reflect the processing capability of the nodes. $r$ is the relay distance and $\lambda_g$ is the gateway density and all BSs associate themselves with their nearest gateways to access to remote server via backhaul. For more details, we recommend the readers to refer to \cite{zhang2016fundamentals}. In short, the summation of the two is the total backhaul delay. 
\end{itemize}

\vspace{-1.50em}
\subsection{Backhaul load per unit area}\vspace{-0.25em}
This performance metric is the backhaul load per unit area. There are total number of $\lambda_u$ users in the unit area. According to association probability that the user is associated with the MCN and SCN with probabilities $p_{am}$ and $p_{a\mu}$, respectively. Instead of considering that the limited backhaul capacity only supports the limited number of users, we characterize the backhaul load by considering the infinite backhaul capacity that supports all cache miss users and the main aim is to study how caching gain (cache hit/content diversity gain) will affect the backhaul offloading with the comparison to the benchmark case.\vspace{-1.25em}
\begin{prop1}
The backhaul load per unit area over all files for full caching generated within a unit time is given as\vspace{-1.75em}
\begin{align}
\mathcal{B} = \sum\nolimits_{i=1}^{F} f_i \Bigg\{(1-p_{m_i}) \lambda_u p_{am} \nu_i + (1-p_{\mu_i}) \lambda_u  (1-p_{am}) \nu_i \Bigg\}
\end{align}
\end{prop1}
\begin{proof}
Since the average number of users in the unit area is $\lambda_u$, the total number of cache miss users requested the $i$th file are $(1-p_{m_i}) \lambda_u  f_i p_{am}$ for mmWave SCN and $(1-p_{\mu_i}) \lambda_\mu  (1-p_{am} f_i$ for $\mu$Wave MCN, respectively. In particular, all backhaul load is generated by all cache miss users with target data rate $\nu_i$. To conclude the proof, we take average over all possible requested files.
\end{proof}
\vspace{-1.50em}
\section{Numerical results}\vspace{-0.75em}
After developing the analytical framework in the previous sections, we now evaluate the performance of the proposed caching placement strategies with respect to two main performance metrics -- retransmission-based ASP of file delivery and average latency with Monte Carlo simulation that verifies our analytical results under the parameter settings given in Table \ref{network settings}. For simplicity, a uniform target rate for each file is considered throughout the analysis. 
\begin{table*}[t!]
\renewcommand{\arraystretch}{0.65}
\centering
\caption{\textbf{Parameter settings}}
\label{network settings}
\begin{tabular}{|l||l|}
\hline
{\textbf{Network design parameter}} & {\textbf{Values}} \\ \hline
{$\lambda_\mu, \lambda_m, \lambda_u, \lambda_g$} & {$5\times10^{-6}, 10^{-5}, 8\times10^{-5}, 5\times10^{-7}$ (nodes/m$^2$)} \\ \hline
{$\mathrm{P}_{\mu}, \mathrm{P}_m$} & {$46, 30$ (dBm)} \\ \hline
{$n_t^\mu, n_t^m, n_r^\mu, n_r^m$} & {$100, 256, 1, 16$} \\ \hline
{$U_\mu = n_t^\mu, U_m=N_{\rm RF}$} & {$100, 10$} \\ \hline
{$W_\mu, W_m$} & {$200$ (MHz), $1$ (GHz)} \\ \hline
{$F$} & {$20$} \\ \hline
{$\nu_i$ ($i = 1, 2, \cdots, F$)} & {$10^{6}$ (bits/s)} \\ \hline
{$\upsilon$} & {$0.8$} \\ \hline
{$C_\mu, C_m$} & {$\{3, 2\}, \{10, 8\}, \{17, 15\}$ (files)} \\ \hline
{$c_1, c_2$} & {$60, 0$} \\ \hline
{$N$} & {$1, 3 , 5$} \\ \hline
{$\beta$} & {$0.008$} \\ \hline
{$\alpha_\LOS, \alpha_\NLOS, \alpha_\mu$} & {$2, 4, 3.5$} \\ \hline
{$B_\mu, B_m$} & {$1, 0.5$} \\ \hline
{$\rho_{\rm UE}, \rho_{\rm BS}$} & {$0.5, 0.5$} \\ \hline
{$\eta_{\LOS}, \eta_{\NLOS}$} & {$3, 5$} \\ \hline
{$r$} & {$200$ (m)} \\ \hline
{$k_1, k_2$} & {$10 , 1$} \\ \hline
{$a, \omega$} & {$10^{-5}$, $10^{-8}$} \\ \hline
{$S$} & {$10^{6}$ (bits)}\\ \hline
\end{tabular}\vspace{-2.5em}
\end{table*}
To obtain further insights on caching gain under different network settings, we begin by evaluating the retransmission-based ASP of file delivery in the hybrid network with respect to various cache size, retransmission attempts, backhaul capacity, the number of files, target data rate, and blockages. In particular, we consider the no caching event as the baseline to compare our results. 

\begin{figure*}[t!]
\centering
\includegraphics[width=1\textwidth]{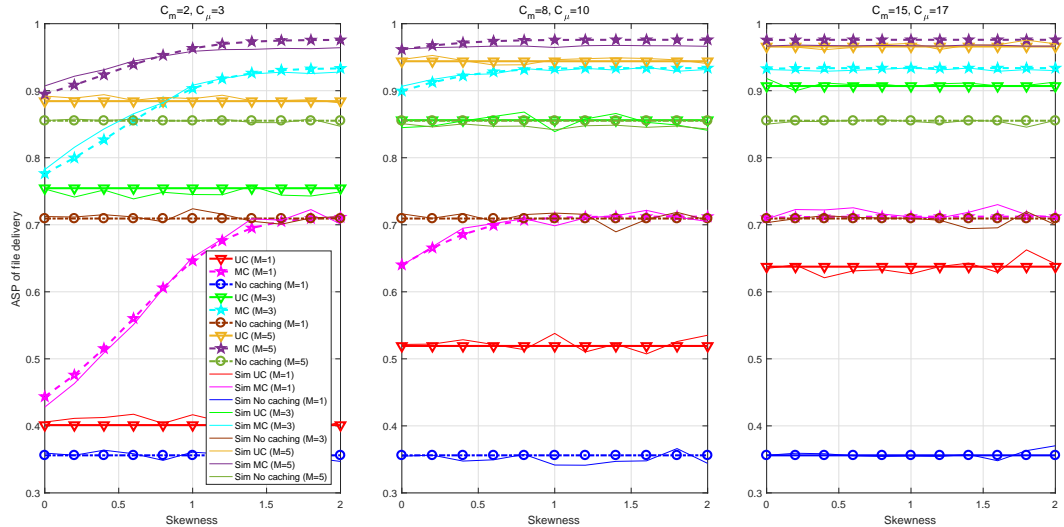}\vspace{-1.0em}
\caption{ASP of file delivery v.s. skewness with various cache size and retransmission attempts}\vspace{-3.50em}
\label{ASP_vs_skewness}
\end{figure*}
In Fig. \ref{ASP_vs_skewness}, given the cache size and retransmission attempt ($N=1$), the performance with caching is better than that without caching for all scenarios throughout the skewness. In particular, the performance of MC is always better than that of UC since MC is sensitive to the content popularity that is characterized by the skewness. When the value of skewness is large, the file index located at the top is more popular than those located at the end; on the contrary, when the value of skewness is smaller, the content popularity tends to be uniform distributed. To conclude, MC is more suitable for higher value of skewness. Now given the cache size and a specific caching scheme, when the retransmission attempts increase, the performance is evidently improved even for no caching scenario. Finally, we evaluate the impact of cache size on the performance. The content diversity gain (\emph{i.e.,} cache hit ratio) is mainly dependent on the cache size. The more cache sizes imply that there are more files can be stored. For example, the caching probability of UC to cache a file will increase as the increase of the cache size. This improvement is shown in Fig. \ref{ASP_vs_skewness}, where the performance is better under higher cache size for a specific caching scheme given the retransmission scheme. In particular, since the cache size at edge nodes is very small, it is necessary to adopt retransmission scheme to further improve the performance under small cache sizes.   

\begin{figure*}
\centering
\includegraphics[width=1\textwidth]{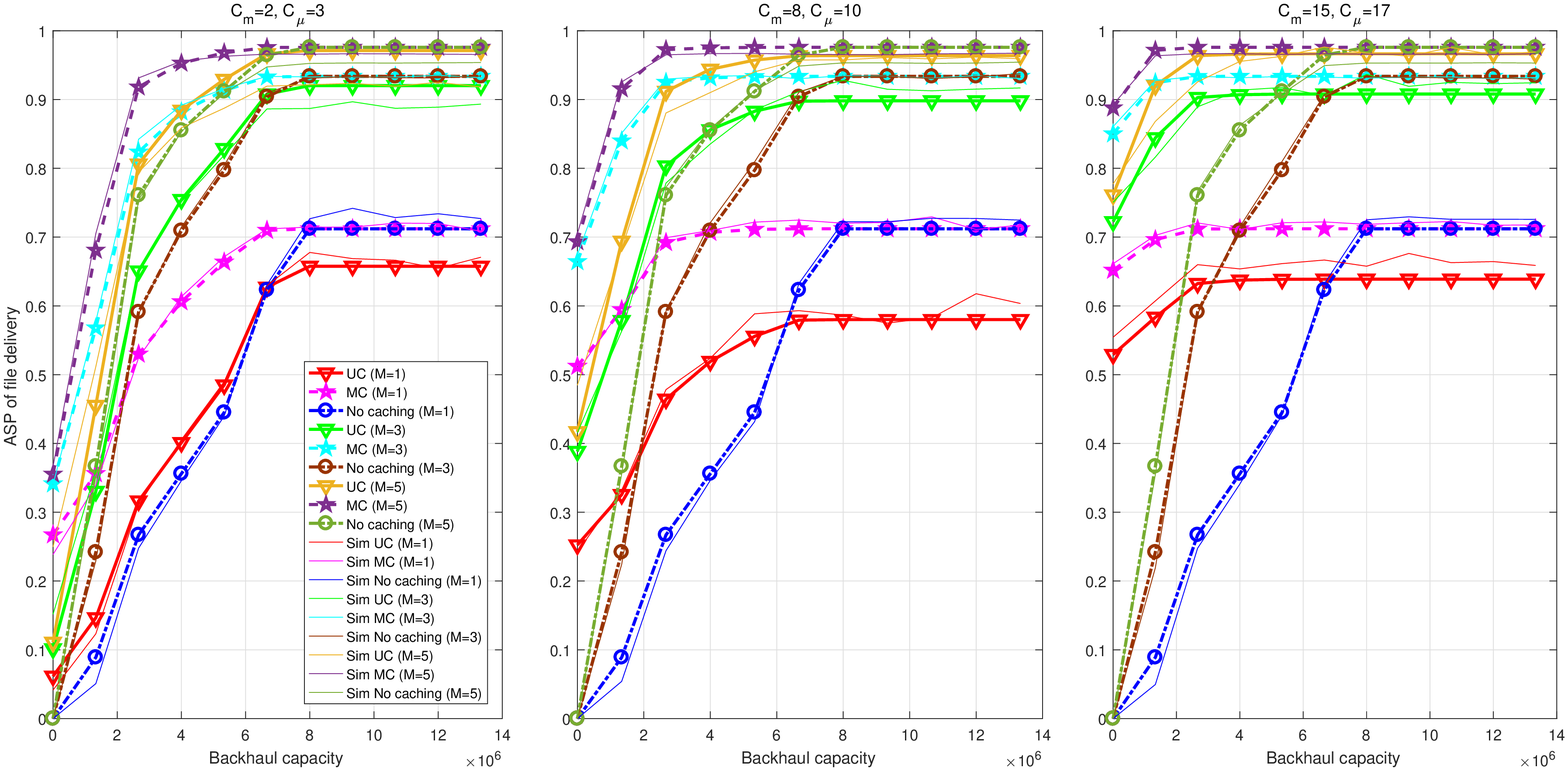}\vspace{-1.0em}
\caption{ASP of file delivery v.s. backhaul capacity with various cache size and retransmission attempts}\vspace{-3.25em}
\label{ASP_vs_backhaul}
\end{figure*}
In Fig. \ref{ASP_vs_backhaul}, we evaluate the performance with respect to the backhaul capacity. Given the cache size and retransmission attempt ($N=1$), the performance with caching is better than that without caching under weak backhaul capacity. And MC is better than UC throughout skewness. Furthermore, as the backhaul capacity tends to be very strong, no cache performance can achieve up to the same as that achieved by MC. However, the performance of UC under strong backhaul capacity cannot achieve up to that achieved by MC. This is due to the fact that UC is not content popularity based caching scheme as MC and any file to be cached or not depends on the caching probabilities that is the function of cache size. In contrast, any file to be stored or not is absolutely determinant for MC. Therefore, for MC, the impact of backhaul capacity on the cache miss part is always certain for fixed cache size while for UC, this impact of backhaul capacity on the cache miss part is dependent to cache miss association probability. When the cache size is very small, namely the caching probability is small and cache miss association probability is large, the impact of backhaul capacity on the performance of UC  has more contribution. To conclude, due to the randomness of caching probabilities for UC, there is a loss gap between UC and MC under strong backhaul capacity. This justification holds for the reason of the increase of the loss gap between UC and MC given more cache size under strong backhaul capacity. When the cache size increases, the performance of weak  backhaul capacity improves but the performance of strong backhaul capacity decreases for UC. Moreover, this demonstrates that retransmission is necessary for weak backhaul capacity to improve the performance.

\begin{figure*}
\centering
\includegraphics[width=1\textwidth]{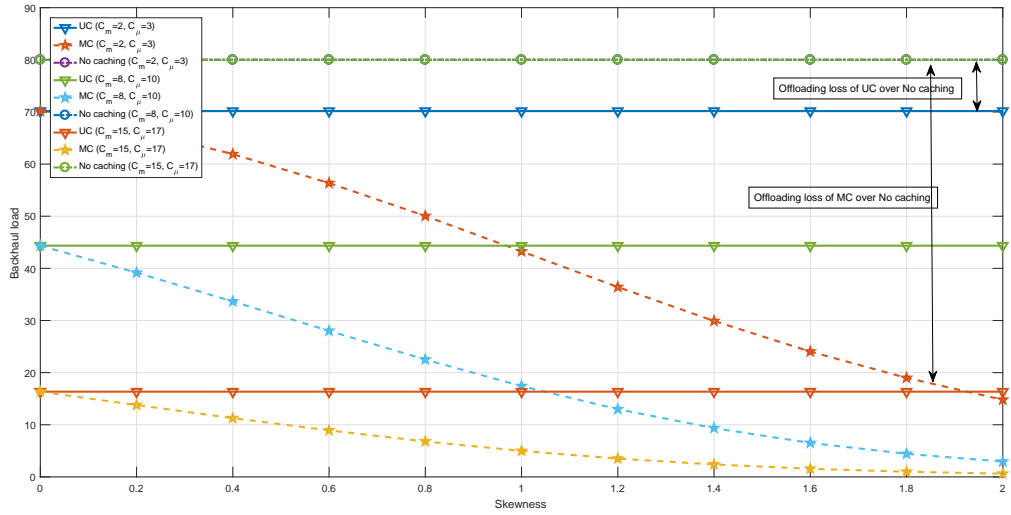}\vspace{-1.0em}
\caption{Backhaul load with various common caching strategies}
\label{backhaul load}\vspace{-1.50em}
\end{figure*}
In particular, as we can see from Fig. \ref{ASP_vs_backhaul}, even though the backhaul capacity is low, the performance is not zero as no caching performance starts from zero. This implies that caching can definitely improve the backhaul load. For better understanding, we also show the backhaul load\footnote{Here, for analysis simplicity, we relax the constraint of backhaul capacity and assume backhaul link is huge enough to always support all cache miss users. This assumption is justifiable since the total backhaul load will finally be upper bounded if considering the backhaul capacity constraints.} in Fig. \ref{backhaul load}. In particular, we show the offloading loss between a specific caching scheme and the no caching base line event. When the cache size increases, the backhaul load decreases and the backhaul offloading gap increases.

\begin{figure*}
\centering
\includegraphics[width=1\textwidth]{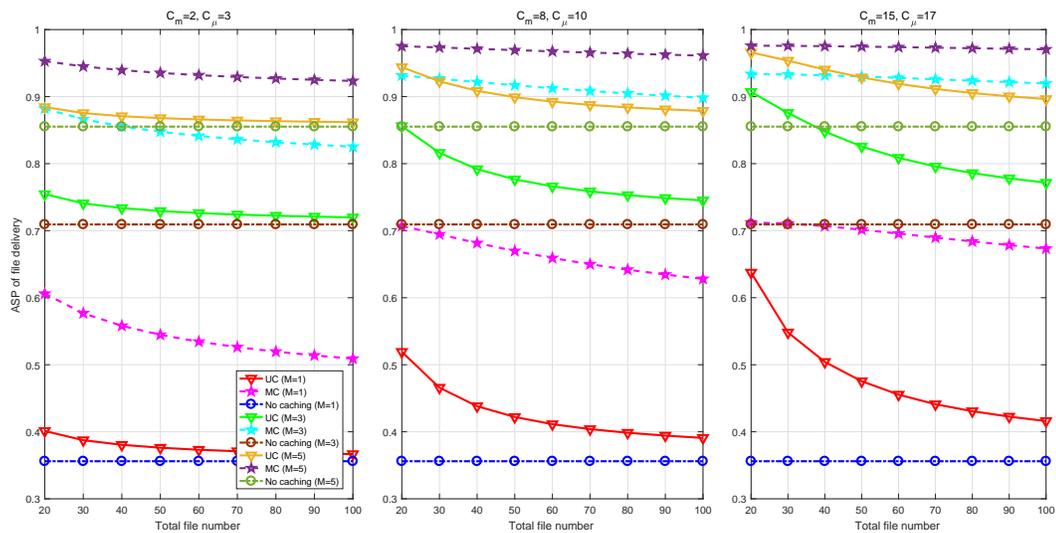}\vspace{-1.0em}
\caption{ASP of file delivery v.s. file number with various cache size and retransmission attempts}\vspace{-4.5em}
\label{ASP_vs_file}
\end{figure*}
In Fig. \ref{ASP_vs_file}, when the file number increases, less number of files can be cached given cache size, which implies that the caching probabilities (or cached files) of UC (or MC) is less. The content diversity gain will automatically fall down. Fig. \ref{ASP_vs_file} demonstrates this phenomenon and the improvement can be made through the increase of cache size and retransmission.

\begin{figure*}
\centering
\includegraphics[width=1\textwidth]{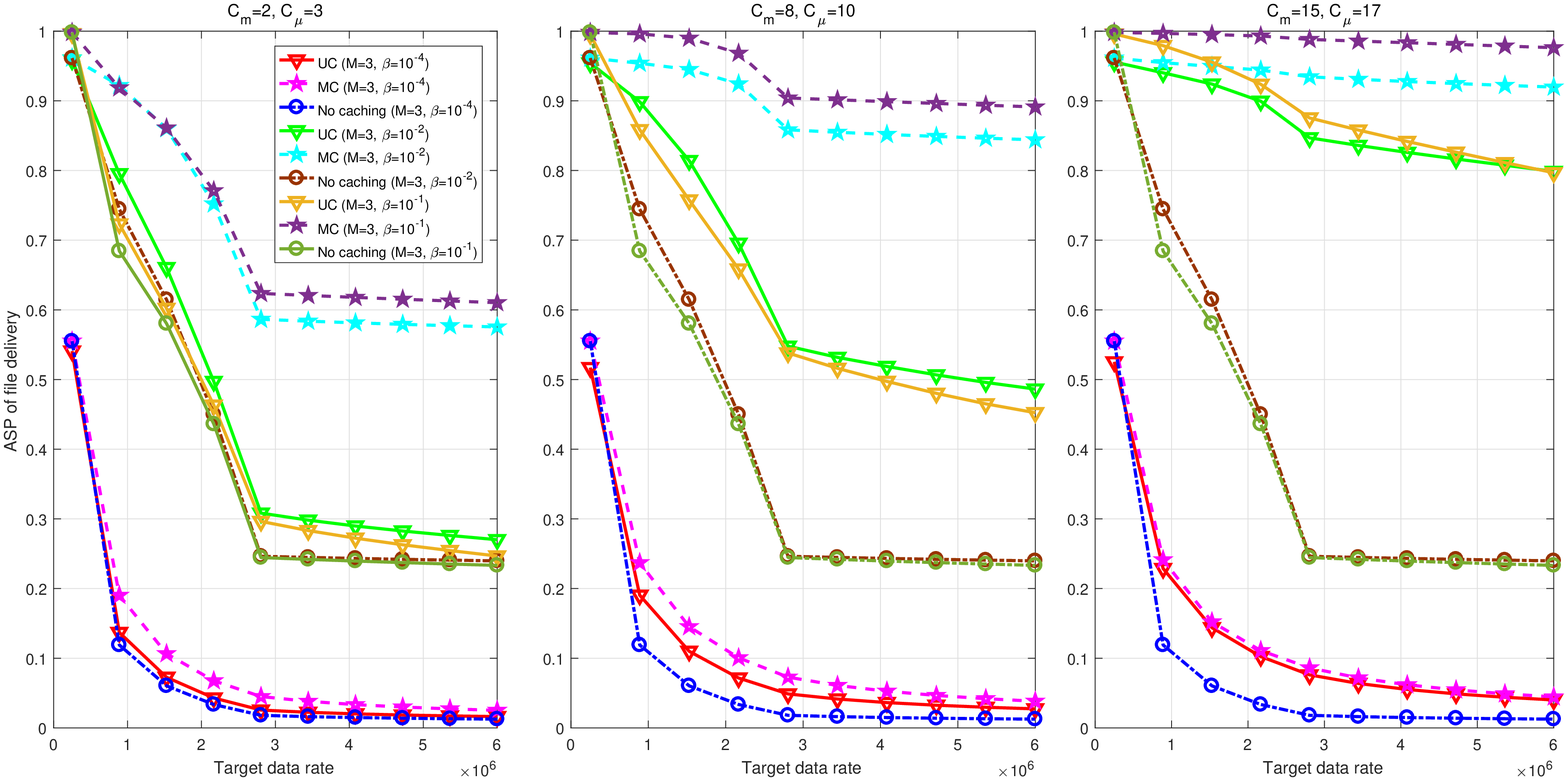}\vspace{-1.0em}
\caption{ASP of file delivery v.s. Target data rate with various cache size and blockage densities}\vspace{-2.50em}
\label{ASP_vs_rate}
\end{figure*} 
\begin{figure*}
\centering
\includegraphics[width=1\textwidth]{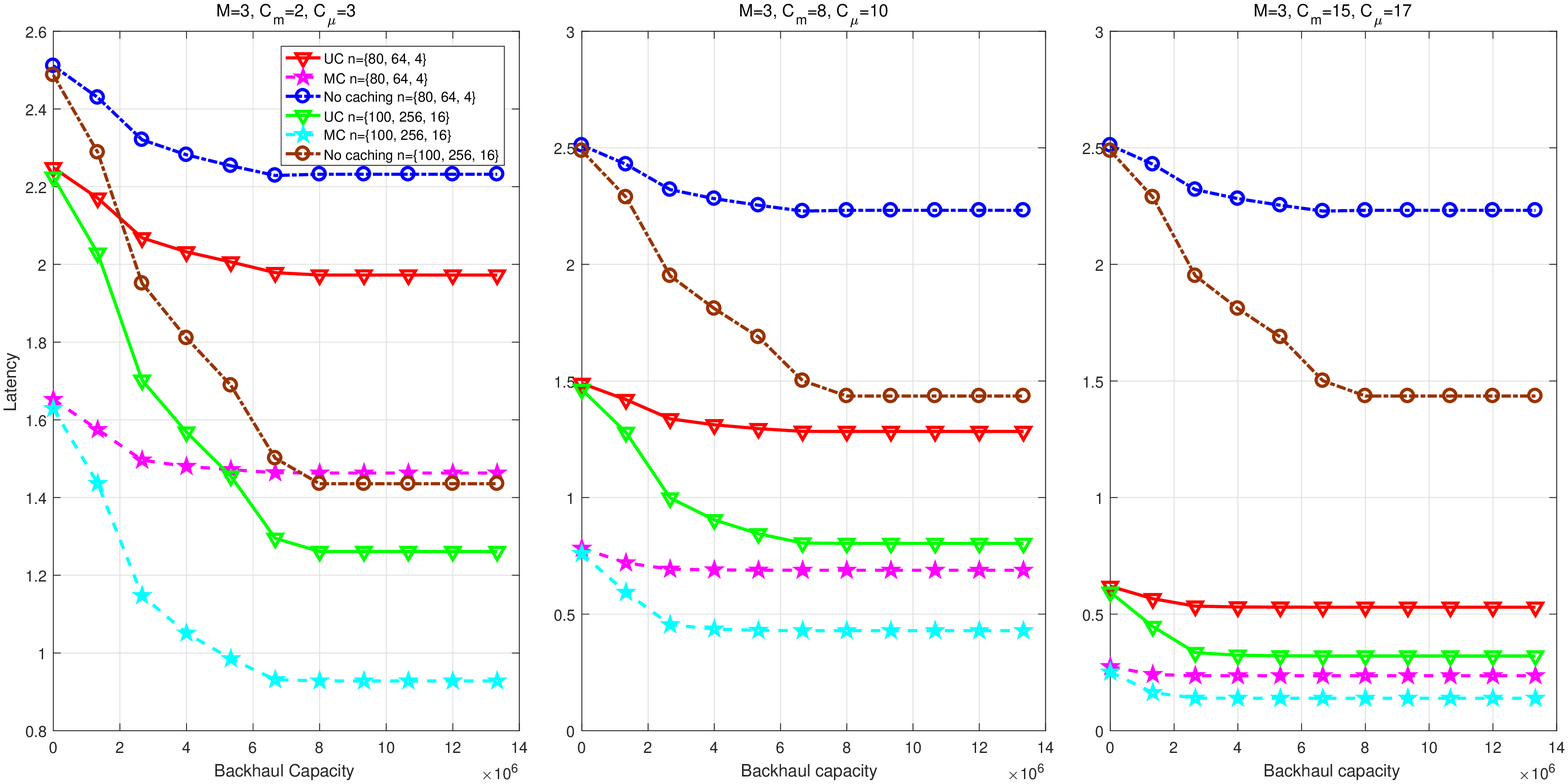}\vspace{-1.0em}
\caption{Latency v.s. backhaul capacity with various cache size and antenna number}\vspace{-1.50em}
\label{Latency_vs_backhaul}
\end{figure*} 
\begin{figure*}
\centering
\includegraphics[width=1\textwidth]{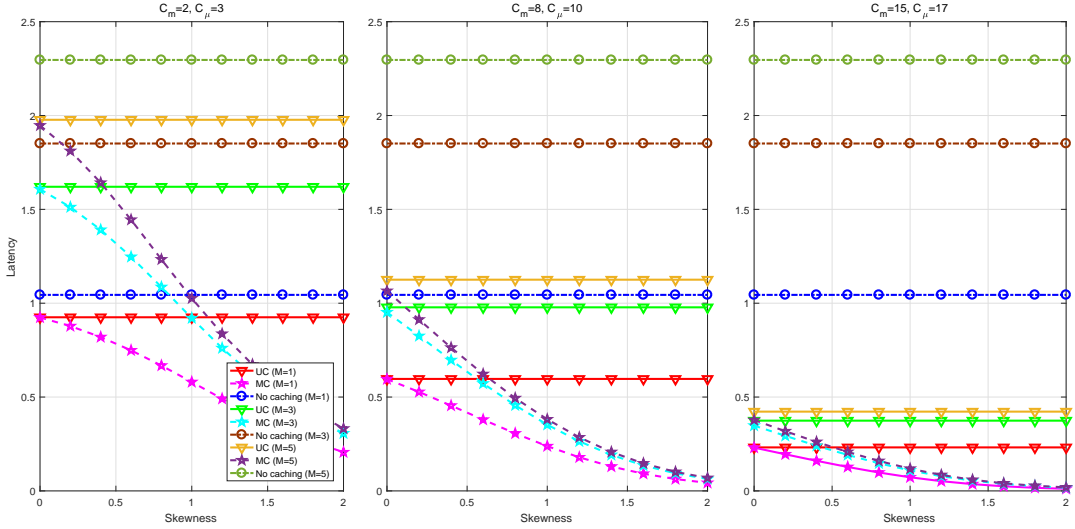}\vspace{-1.0em}
\caption{Latency v.s. skewness with various cache size and retransmission attempt}\vspace{-1.50em}
\label{Latency_vs_skewness}
\end{figure*} 
Finally, for ASP of file delivery performance, we evaluate another two fundamental network design parameters -- target rate and blockage densities given retransmission. Given the cache size, when the blockage density increases, the performance improves since $1)$ mmWave network under higher blockage density is more noise-limited and SINR will be higher, $2)$ mmWave NLOS and $\mu$Wave association probability increase and non-blockage-sensitive $\mu$Wave network makes sure more users can be served. 

Now we evaluate transmission latency with respect to backhaul capacity and skewness. In Fig. \ref{Latency_vs_backhaul}, we show that more antenna number will increase the gap of the latency between any caching scheme and no caching event. When the cache size increases, the latency decreases since backhaul processing time can be avoided.    

In Fig. \ref{Latency_vs_skewness}, MC performs better for higher value of skewness than lower value of skewness. Therefore, the latency for MC decreases along skewness axis. When the retransmission attempts increase, the latency increases as a whole but the gap between the MC and UC and MC and no caching event increases. There is a tradeoff between the performance and retransmission. When the retransmission number increases, the ASP of file delivery improves but latency decreases.

\appendices
\setcounter{equation}{0}
\renewcommand{\theequation}{A.\arabic{equation}}\vspace{-1.50em}
\section{Proof of Lemma \ref{association mmWave}}\label{Appears_A}
According to the  least biased path loss association criterion, the association event that the typical user requesting the $i$th file from the file set $\mathcal{F}$ associated with a mmWave BS located at $x_{m_i}^\LOS$ caching the requested file in LOS transmission means that the biased path loss at the typical user from the mmWave BS $x_{m_i}^\LOS$ is lower than other five cases, which can be formulated as follows\vspace{-1.0em}
\begin{align}
p_{x^\LOS_{m_i}}=&\mathbb{E}_{r_{x^\LOS_{m_i}}}\Big\{\mathbb{P}[B_m r_{x^\LOS_{m_i}}^{-\alpha_\LOS} \geq B_\mu r_{x_{\mu_i}}^{-\alpha_\mu}|r_{x^\LOS_{m_i}}=R]\times\mathbb{P}[B_m r_{x^\LOS_{m_i}}^{-\alpha_\LOS} \geq B_\mu r_{y_{\mu_i}}^{-\alpha_\mu}|r_{x^\LOS_{m_i}}=R]\nonumber \\
&\qquad\;\;\times\mathbb{P}[B_m r_{x^\LOS_{m_i}}^{-\alpha_\LOS} \geq B_m  r^{-\alpha_\NLOS}_{x^\NLOS_{m_i}}|r_{x^\LOS_{m_i}}=R]\,\,\mathbb{P}[B_m r_{x^\LOS_{m_i}}^{-\alpha_\LOS} \geq B_m  r^{-\alpha_\LOS}_{y^\LOS_{m_i}}|r_{x^\LOS_{m_i}}=R]\nonumber \\
&\qquad\;\;\times\mathbb{P}[B_m r_{x^\LOS_{m_i}}^{-\alpha_\LOS} \geq B_m  r^{-\alpha_\NLOS}_{y^\NLOS_{m_i}}|r_{x^\LOS_{m_i}}=R]\Big\}\nonumber \\
=&\int_{0}^{\infty} p_{1}\,p_{2}\,p_{3}\,p_{4}\,p_{5}\,f_{r_{x^\LOS_{m_i}}}(R) \d R\,,\label{Association probability user to mmwave LOS}
\end{align}
By using the result of void probability, the first term $p_1$ can be calculated as \vspace{-1.0em}
\begin{align}
p_{1} &= \mathbb{P}[B_m r_{x^\LOS_{m_i}}^{-\alpha_\LOS} \geq B_\mu r_{x_{\mu_i}}^{-\alpha_\mu}|r_{x^\LOS_{m_i}}=R]\nonumber \\
&=\mathbb{P}[r_{x_{\mu_i}} \geq (\frac{B_\mu R^{\alpha_\LOS}}{B_m})^{\frac{1}{\alpha_\mu}}]\nonumber \\
&=\mathbb{P}\Big[\Phi_{\mu_i}\Big([0,(\frac{B_\mu R^{\alpha_\LOS}}{B_m})^{\frac{1}{\alpha_\mu}}]\Big) = 0 \Big]\nonumber \\
&=\exp\Big[-\pi \lambda_\mu p_{\mu_i} (\frac{B_\mu R^{\alpha_\LOS}}{B_m})^{\frac{2}{\alpha_\mu}}\Big]\,.\label{first term association probability}
\end{align}
Similarly, all other terms can be calculated by following the same derivation. Finally, the distribution of the distance $R$ between any two points from a PPP follows from the nearest neighbour distance distribution given by\vspace{-1.0em}
\begin{align}\label{nearest neighbour distance distribution}
f_{r_{x_{m_i}^\LOS}}(R) =  2 \pi \lambda_{\mu} p_{\mu_i} R \exp(-\pi \lambda_\mu p_{\mu_i} R^2 )\,.
\end{align}
By substituting \eqref{first term association probability} and \eqref{nearest neighbour distance distribution} into \eqref{Association probability user to mmwave LOS}, the proof is concluded. In the similar way, other association probabilities are calculated.
\vspace{-1.0em}
\setcounter{equation}{0}
\renewcommand{\theequation}{B.\arabic{equation}}\vspace{-1.0em}
\section{Proof of Lemma \ref{Distribution of distance between serving BS and typical user}}
\label{Appears_B}
Assume the typical requesting the $i$th file served by a mmWave BS located at $x_{m_i}^\LOS$ with LOS transmission and the distance between them is $R_{x_{m_i}^\LOS}$. This proof shows how the PDF of $R_{x_{m_i}^\LOS}$ is computed and other distances are similarly proofed.

At first, we consider the event $R_{x_{m_i}^\LOS} > D$ is equivalent to that $R_{x_{m_i}^\LOS} > D$ given that the typical user is associated with the serving mmWave SBS located at $x_{m_i}^\LOS$ \emph{i.e.,}\vspace{-1.0em}
\begin{align}
\mathbb{P}[R_{x_{m_i}^\LOS} > D] &= \mathbb{P}[R_{x_{m_i}^\LOS} > D | \text{The serving BS is } x_{m_i}^\LOS] \nonumber \\
&=\frac{\mathbb{P}[R_{x_{m_i}^\LOS} > D, \text{The serving BS is } x_{m_i}^\LOS]}{\mathbb{P}[\text{The serving BS is } x_{m_i}^\LOS]}\,.\label{PDF of the distance}
\end{align}
The numerator is given as similar calculation as in \eqref{Association probability user to mmwave LOS} but slightly change the lower limit from $0$ to $D$. The denominator is given as $p_{x_{m_i}^\LOS}$. Finally, by substituting the numerator and denominator in the \eqref{PDF of the distance}, the proof is concluded.
\vspace{-1.5em}
\setcounter{equation}{0}
\renewcommand{\theequation}{C.\arabic{equation}}
\section{Proof of Proposition \ref{ASP of file delivery mm}}
\label{Appears_C}
This proof gives the derivation of the conditional ASP of file delivery that the typical user requesting the $i$th file served by the mmWave BS located at $x_{m_i}^\LOS$ with LOS transmission. The derivation of others in the Proposition \ref{ASP of file delivery mm} are the similar. According to the definition of the ASP of file delivery and SINR expression in \eqref{sinr mm approximation}, we have\vspace{-1.0em}
\begin{align}
\mathbb{P}\Big[\frac{W_m}{U_m} \text{log}(1 +\text{SINR}_{x_{m_i}^\LOS})\geq \nu_i\Big]
&\approx\mathbb{P}\Big[\frac{\frac{\mathrm{P}_m}{{U}_{m}}\frac{n^m_r n^m_t}{\eta_{\LOS}}\mathcal{X}_{x_{m_i}^{\LOS}}^2 R_{x_{m_i}^{\LOS}}^{-\alpha_\LOS} p_{ZF}}{\sigma_m^2 + I_{x_{m_i}^{\LOS}}} \geq Q_i\Big]\nonumber \\
&=(1 - \frac{1}{n^m_r})^{({U}_{m} - 1)} \mathbb{E}_{R_{x_{m_i}^\LOS}, \mathcal{X}_{x_{m_i}^{\LOS}}^2, I_{x_{m_i}^{\LOS}}}\Big\{\mathbb{P}\Big[\mathcal{X}_{x_{m_i}^\LOS}^2 \geq \frac{Q_i(\sigma^2_m + I_{x_{m_i}^{\LOS}})}{G_{x_{m_i}^\LOS} R_{x_{m_i}^{\LOS}}^{-\alpha_{\LOS}}}\Big]\Big\}\,,
\end{align}
where $G_{x_{m_i}^\LOS} = \frac{\mathrm{P}_m}{{U}_{m}}\frac{n^m_r n^m_t}{\eta_{\LOS}}$ and $Q_i = 2^{\frac{\nu_i U_m}{W_m}} - 1$. Now we focus on the expectation term and compute the expectation below. \vspace{-0.75em}
\begin{align}
\mathbb{E}\Big\{\mathbb{P}\Big[\mathcal{X}_{x_{m_i}^\LOS}^2 \geq \frac{Q_i(\sigma^2_m + I_{x_{m_i}^\LOS})}{G_{x_{m_i}^\LOS} R_{x_{m_i}^\LOS}^{-\alpha_\LOS}}\Big]\Big\}&=\int_{0}^{\infty} \mathbb{E}\Big\{\mathbb{P}[\mathcal{X}_{x_{m_i}^\LOS}^2 \geq {\frac{Q_i(\sigma^2_m + I_{x_{m_i}^\LOS})}{G_{x_{m_i}^\LOS} R_{x_{m_i}^\LOS}^{-\alpha_L}}}]\Big\}f(R_{x_{m_i}^\LOS})\d R_{x_{m_i}^\LOS}\nonumber \\
&\overset{(a)}{=}\int_{0}^{\infty} \mathbb{E}_{I_{x_{m_i}^\LOS}} \Big\{ \exp\Big(-\frac{Q_i(\sigma^2_m + I_{x_{m_i}^\LOS})}{G_{x_{m_i}^\LOS} R_{x_{m_i}^\LOS}^{-\alpha_L}}\Big)\Big\}f(R_{x_{m_i}^\LOS})\d R_{x_{m_i}^\LOS}\nonumber \\
&=\int_{0}^{\infty} \exp\Big(\frac{-Q_i \sigma^2_m }{G_{x_{m_i}^\LOS} R_{x_{m_i}^\LOS}^{-\alpha_{\LOS}}}\Big) \mathbb{E}_{I_{x_{m_i}^\LOS}}\Big\{\exp\Big(\frac{-Q_i I_{x_{m_i}^\LOS}}{G_{x_{m_i}^\LOS} R_{x_{m_i}^\LOS}^{-\alpha_L}}\Big)\Big\}f(R_{x_{m_i}^\LOS})\d R_{x_{m_i}^\LOS}\,,\label{PG}
\end{align}
where the distribution of the distance between the serving BS and the typical user is given as \eqref{distance LOS cache hit}. $(a)$ follows from the exponential distribution and $f(R_{x_{m_i}^\LOS})$ is provided in Lemma \ref{Distribution of distance between serving BS and typical user}. Now the aim is to compute the expectation term in the integral regarding to interference. However, due to the thinning theorem, the interference can be partitioned into four terms, namely $I_{x_{m_i}^\LOS} = I_{\Phi_{m_i}^\LOS\backslash\{x_{m_i}^\LOS\}} + I_{\Phi_{m_i}^\NLOS} + I_{\overline{\Phi}_{m_i}^\LOS} + I_{\overline{\Phi}_{m_i}^\NLOS}$, where the thinned PPPs $\Phi_{m_i}$ and $\bar\Phi_{m_i}$ regarding to cache hit and cache miss are further thinned regarding to LOS and NLOS transmissions. Therefore, the expectation term in \eqref{PG} is reduced to\vspace{-1.0em}
\begin{align}
\mathbb{E}_{I_{x_{m_i}^\LOS}}\Big\{\exp\Big(\frac{-Q_iI_{x_{m_i}^\LOS}}{G_{x_{m_i}^\LOS} R_{x_{m_i}^\LOS}^{-\alpha_\LOS}}\Big)\Big\}=&\mathbb{E}_{I_{\Phi_{m_i}^\LOS/\{x_{m_i}^\LOS\}}}\Big\{\frac{-Q_iI_{\Phi_{m_i}^\LOS}}{G_{x_{m_i}^\LOS} R_{x_{m_i}^\LOS}^{-\alpha_\LOS}}\Big\} \mathbb{E}_{I_{\Phi_{m_i}^\NLOS}}\Big\{\frac{-Q_iI_{\Phi_{m_i}^\NLOS}}{G_{x_{m_i}^\LOS} R_{x_{m_i}^\LOS}^{-\alpha_\LOS}}\Big\} \nonumber \\
&\mathbb{E}_{I_{\overline{\Phi}_{m_i}^\LOS}}\Big\{\frac{-Q_iI_{\overline{\Phi}_{m_i}^\LOS}}{G_{x_{m_i}^\LOS} R_{x_{m_i}^\LOS}^{-\alpha_\LOS}}\Big\} \mathbb{E}_{I_{\overline{\Phi}_{m_i}^\NLOS}}\Big\{\frac{-Q_iI_{\overline{\Phi}_{m_i}^\NLOS}}{G_{x_{m_i}^\LOS} R_{x_{m_i}^\LOS}^{-\alpha_\LOS}}\Big\}\,.
\end{align}
Now we take the first term as the example to show how to compute the expectation and all other terms are computed in a similar way. For analytical tractability, we give the upper and lower bound closed-form, respectively.
Assume that $s = \frac{-Q_i}{G_{x_{m_i}^\LOS}R_{x_{m_i}^\LOS}^{-\alpha_\LOS}}$, then \vspace{-1.0em}
\begin{align}
&\mathbb{E}_{I_{\Phi_{m_i}^\LOS\backslash\{x_{m_i}^\LOS\}}}\Big\{\exp\Big(\frac{-Q_iI_{\Phi_{m_i}^\LOS}}{G_{x_{m_i}^\LOS} R_{x_{m_i}^\LOS}^{-\alpha_\LOS}}\Big)\Big\} 
=\mathbb{E}\Big\{\exp\Big(s I_{\Phi_{m_i}^\LOS} \Big)\Big\}\nonumber \\
&{=}\mathbb{E}\left\{\exp\Big(s \sum\nolimits_{\substack{y \in \Phi_{m_i}^\LOS,\;y\neq x_{m_i}^\LOS}} \frac{\mathrm{P}_m}{{U}_{y}} \frac{n^m_r n^m_t}{\eta_{\LOS}} r_{y}^{-\alpha_\LOS}  \sum\nolimits_{u \in \mathcal{U}_{y}}  ||\sum\nolimits_{k=1}^{\eta_{\LOS}}{\mathcal{X}_{k,y}}\gamma_{y,u}||^2\Big)\right\}\nonumber \\
&\overset{(a)}{\leq}\mathbb{E}\Big\{\prod_{\substack{y\in \Phi_{m_i}^{\LOS},\;y\neq x_{m_i}^{\LOS}}}\mathbb{E}\Big\{\exp\Big(s \mathrm{P}_m \frac{n^m_r n^m_t}{\eta_{\LOS}} r_{y}^{-\alpha_\LOS}  (\rho_{BS}\rho_{UE})^2 {||\sum_{k=1}^{\eta_{\LOS}}{\mathcal{X}_{k,y}}||^2}\Big)\Big\}\Big\}\nonumber \\
&\overset{(b)}{=}\mathbb{E}\Big\{\prod_{\substack{y\in \Phi_{m_i}^\LOS,\;y\neq x_{m_i}^\LOS}}\Big({1 - {s \mathrm{P}_m n^m_r n^m_t r_{y}^{-\alpha_\LOS} (\rho_{BS}\rho_{UE})^2}}\Big)^{-1}\Big\}\nonumber\\
&\overset{(c)}{=}\exp\Big\{-\int_{R}^{\infty} \Big[1 - \Big(\frac{1}{1 - {s \mathrm{P}_m n^m_r n^m_t r^{-\alpha_\LOS} (\rho_{BS}\rho_{UE})^2}}\Big)\Big]2 \pi \lambda_m p_\LOS(r) r p_{m_i} \d r\Big\}\,,
\end{align} 
where $(a)$ follows from the fact that $\gamma_{y,u} \leq \rho_{\rm UE} \rho_{\rm BS}$ as mentioned in \eqref{inter}. $(b)$ follows from the fact that $||\sum_{k=1}^{\eta_\LOS} \mathcal{X}_{k,y}||^2 \sim \rm Exp(\frac{1}{\eta_\LOS})$ and then the Laplace transform of the channel power gain random variable gives the result. $(c)$ follows from the probability generating functional of a PPP.

However, if we take $\rho_{\rm UE} = \rho_{\rm BS} = 1$ to acquire the lower bound, it is not much tight unless the path loss from the interfering mmWave SBSs is only one. Therefore, in the following we give the tight lower bound by using Cauchy-Schwarz inequality.\vspace{-1.0em}
\begin{align}
&\mathbb{E}_{I_{\Phi_{m_i}^\LOS\backslash\{x_{m_i}^\LOS\}}}\Big\{\exp\Big(\frac{-Q_iI_{\Phi_{m_i}^\LOS}}{G_{x_{m_i}^\LOS} R^{-\alpha_\LOS}_{x_{m_i}^\LOS}}\Big)\Big\}=\mathbb{E}\Big\{\exp\Big(s I_{\Phi_{m_i}^\LOS} \Big)\Big\}\nonumber \\
&{=}\mathbb{E}\left\{\exp\Big(s \sum_{\substack{y\in \Phi_{m_i}^\LOS,\;y\neq x_{m_i}^\LOS}} \frac{\mathrm{P}_m}{{U}_{y}} \frac{n^m_r n^m_t}{\eta_{\LOS}} r_{y}^{-\alpha_\LOS}  \sum_{u \in \mathcal{U}_{y}}  ||\sum_{k=1}^{\eta_{\LOS}}{\mathcal{X}_{k,y}}\gamma_{y,u}||^2\Big)\right\}\nonumber \\
&\overset{(a)}{\geq}\mathbb{E}\Big\{\prod_{\substack{y\in \Phi_{m_i}^{\LOS},\;y\neq x_{m_i}^{\LOS}}}\mathbb{E}\Big\{\exp\Big(s \frac{\mathrm{P}_m}{{U}_{y}} \frac{n^m_r n^m_t}{\eta_{\LOS}} r_{y}^{-\alpha_\LOS} (\sum_{u \in \mathcal{U}_y} \sum_{k=1}^{\eta_{\LOS}} \gamma_{y,u}^2) {\sum_{k=1}^{\eta_{\LOS}}||{\mathcal{X}_{k,y}}||^2}\Big)\Big\}\Big\}\nonumber \\
&\overset{(b)}{=}\mathbb{E}\Big\{\prod_{\substack{y\in \Phi_{m_i}^\LOS,\;y\neq x_{m_i}^\LOS}}\Big({1 - {s \frac{\mathrm{P}_m}{{U}_y} \frac{n^m_r n^m_t}{\eta_{\LOS}} r_{y}^{-\alpha_\LOS} (\sum_{u \in \mathcal{U}_y} \sum_{k=1}^{\eta_{\LOS}} \gamma_{y,u}^2 }})\Big)^{-\eta_{\LOS}}\Big\}\nonumber\\
&\overset{(c)}{\geq}\exp\Big\{-\int_{R}^{\infty} \Big[1 - \Big(\frac{1}{1 - {s \mathrm{P}_m n^m_r n^m_t r^{-\alpha_\LOS} }}\Big)^{\eta_{\LOS}}\Big]2 \pi \lambda_m p_{\LOS}(r) r p_{m_i} \d r\Big\}\,,
\end{align}
where $(a)$ follows from the Cauchy-Schwarz inequality. $(b)$ follows from the fact that ${\sum_{k=1}^{\eta_{\LOS}}||\sqrt{\mathcal{X}_{k,y,0}}||^2}$ follows Chi-squre/gamma distribution with parameters $\eta_{\LOS}$ and 1. $(c)$ follows from that $\gamma_{y,u}^2 \leq 1$ and the probability generating functional of a PPP.

In the similar manner, the other expectation terms are upper bounded and lower bounded but with different integral lower limits. Finally, substituting these intermediate results into \eqref{PG}, the proof is compete.
\vspace{-1.0em}
\setcounter{equation}{0}
\renewcommand{\theequation}{D.\arabic{equation}}\vspace{-1.50em}
\section{Proof of Lemma \ref{achievable data rate mu}}
\label{Appears_D}
According to \cite{6816003} Lemma 1, when the number of antennas is large, we have the approximation provided below.\vspace{-2.50em}
\begin{align}
\mathbb{E} \Big\{ \rm{log}_2\Big(1 + \frac{X}{Y}\Big)\Big\} \approx\text{log}_2\Big(1 + \frac{\mathbb{E}(X)}{\mathbb{E}(Y)}\Big).
\end{align}
The achievable data rate $\mathcal{\bar R}_{x_{\mu}}$ is given by\vspace{-1.0em}
\begin{align}
\mathcal{\bar R}_{x_{\mu_i}} &= \mathbb{E}\Big\{\frac{W_\mu}{U_\mu} \text{log}(1 + \text{SINR}^{\rm ZF}_{x_{\mu_i}})\Big\}=\frac{W_\mu}{U_\mu} \mathbb{E}\Big\{\text{log}(1 + \text{SINR}^{\rm ZF}_{x_{\mu_i}})\Big\}\nonumber \\
&\approx \frac{W_\mu}{U_\mu} \text{log}\left(1 + \frac{\mathbb{E}\Big\{\frac{\mathrm{P}_\mu}{{U}_{\mu}}(\mathbb{E}[\sqrt{\mathcal{G}_{x_{\mu_i}}}])^2 r_{x_{\mu_i}}^{-\alpha_\mu}\Big\}}{\mathbb{E}\Big\{\frac{\mathrm{P}_\mu}{{U}_{\mu}}(\sqrt{\mathcal{G}_{x_{\mu_i}}}-\mathbb{E}[\sqrt{\mathcal{G}_{x_{\mu_i}}}] )^2 r_{x_{\mu_i}}^{-\alpha_\mu} + \sum_{\substack{k\in\Phi_{\mu},\;k\ne x_{\mu_i}}} \frac{\mathrm{P}_\mu}{{U}_{k}} \mathcal{G}_k r_k^{-\alpha_\mu} + \sigma_\mu^2\Big\}}\right)\nonumber \\
&=\frac{W_\mu}{U_\mu} \text{log}\left( 1 + \frac{C_1 r_{x_{\mu_i}}^{-\alpha_\mu}}{C_2 r_{x_{\mu_i}}^{-\alpha_\mu} + C_3 + \sigma_\mu^2}\right)\,
\end{align}
where $C_1 = \frac{\mathrm{P}_\mu}{{U}_{\mu}}(\mathbb{E}[\sqrt{\mathcal{G}_{x_{\mu_i}}}])^2$. $C_2 = \mathbb{E}\Big\{\frac{\mathrm{P}_\mu}{{U}_{{\mu}}}(\sqrt{\mathcal{G}_{x_{\mu_i}}}-\mathbb{E}[\sqrt{\mathcal{G}_{x_{\mu_i}}}] )^2\Big\}$. $C_3 = \mathbb{E}\Big\{\sum_{\substack{k\in\Phi_{\mu},\;k\ne x_{\mu_i}}} \frac{\mathrm{P}_\mu}{{U}_{k}} \mathcal{G}_k r_k^{-\alpha_\mu}\Big\}$. Due to the fact that $\mathcal{G_{x_{\mu_i}}} \sim \Gamma(n^\mu_t - U_\mu + 1, 1 )$ and $\mathcal{G}_k \sim \Gamma(U_k,1)$, we have\vspace{-0.5em}
\begin{align}
C_1 &= \frac{\mathrm{P}_\mu}{{U}_{\mu}}\Big(\frac{\Gamma(n_t^\mu - {U}_{\mu} + \frac{3}{2})}{\Gamma(n_t^\mu - {U}_{\mu} + 1)}\Big)^2,\;&C_2& = \frac{\mathrm{P}_\mu}{\mathcal{U}_{x_{\mu}}} \Big(n_t^\mu - {U}_{\mu} + 1\Big) - C_1,
\end{align}
\begin{align}
C_3 &= \mathbb{E}\Big\{\sum_{\substack{k\in\Phi_{\mu},\;k\ne x_{\mu_i}}} \frac{\mathrm{P}_\mu}{{U}_{k}} \mathcal{G}_k r_k^{-\alpha_\mu}\Big\}{=}\mathbb{E}_{\Phi_{\mu}}\{\sum_{\substack{k \in \Phi_{\mu_i},\;k \neq x_{\mu_i}}}\mathrm{P}_\mu r_k^{-\alpha_\mu} + \sum_{k' \in \bar \Phi_{\mu_i}} \mathrm{P}_\mu r_{k'}^{-\alpha_\mu}\}\nonumber \\
&\overset{(a)}{=}\int_{R_{x_{\mu_i}}}^{\infty} \mathrm{P}_\mu r^{-\alpha_\mu} 2 \pi p_{\mu_i} \lambda_\mu  r \d r + \int_{1}^{\infty} \mathrm{P}_\mu r^{-\alpha_\mu} 2 \pi (1-p_{\mu_i}) \lambda_\mu  r \d r \nonumber \\
&= \underbrace{\mathrm{P}_\mu 2 \pi p_{\mu_i} \lambda_\mu  \frac{1}{\alpha_\mu - 2}}_{C_3^{'}} r_{x_{\mu}}^{-\alpha_\mu + 2} + \underbrace{\mathrm{P}_\mu 2 \pi (1 - p_{\mu_i} ) \lambda_\mu \frac{1}{\alpha_\mu - 2}}_{C_3^{''}}\,,
\end{align}
where $(a)$ follows from the Campell's theorem. Likewise, the other achievable data rate of the cache miss scenario is acquired.
\vspace{-4.50em}
\setcounter{equation}{0}
\renewcommand{\theequation}{E.\arabic{equation}}\vspace{2.5em}
\section{Proof of Proposition \ref{Prop_conditional_ASP_mu}}
\label{Appears_E}
By using the stochastic geometric analysis, we have\vspace{-1.0em}
\begin{align}
\mathbb{P}[\mathcal{\bar R}_{x_{\mu_i}} \geq \nu_i]&=\mathbb{P}\Big[\frac{W_\mu}{U_\mu} \text{log}\Big(1 + \frac{C_1 R_{x_{\mu_i}}^{-\alpha_\mu}}{C_2 R_{x_{\mu_i}}^{-\alpha_\mu} + C'_3 R_{x_{\mu_i}}^{-\alpha_\mu+2}  +C''_3+ \sigma_\mu^2}\Big) \geq \nu_i\Big]\nonumber \\
& = \mathbb{P}\Big[\frac{C_1 R_{x_{\mu_i}}^{-\alpha_\mu}}{C_2 R_{x_{\mu_i}}^{-\alpha_\mu} + C'_3 R_{x_{\mu_i}}^{-\alpha_\mu+2}  +C''_3+ \sigma_\mu^2} \geq \underbrace{ 2^{\frac{U_\mu \nu_i}{W_\mu}} - 1}_{\hat Q_{i}}\Big]\nonumber \\
&=\mathbb{P}\Big[\sigma_\mu^2 \hat Q_{i} R_{x_{\mu}}^{\alpha_\mu} + C'_3 \hat Q_{i} R_{x_{\mu}}^{2} \leq C_1 - C_2 \hat Q_{i}\Big]\,,
\end{align}
It is worth noting that $R^*$ is the numerical solution to the function $C''_3 R_{x_{\mu_i}}^{\alpha_\mu} \hat Q_{i} + \sigma_\mu^2 \hat Q_{i} R_{x_{\mu_i}}^{\alpha_\mu} + C'_3 Q_{\mu_i} R_{x_{\mu_i}}^{2} \leq C_1 - C_2 \hat Q_{i}$. To compute the function is not easy, we numerically solve the function and give the lower bound  i.e., $\floor{R^*}$. Therefore, the conditional ASP of file delivery is \vspace{-1.0em}
\begin{align}
\mathbb{P}[\mathcal{\bar R}_{x_{\mu_i}} \geq \nu_i]&\geq \mathbb{P}\Big[R_{x_{\mu_i}} \leq \floor{R^*}\Big]\nonumber \\
&=\int_{1}^{\floor{R^*}} f(R_{x_{\mu_i}}) \textup{d} R_{x_{\mu_i}}\,,
\end{align}
The upper bounded conditional ASP of file delivery is also given as\vspace{-1.0em}
\begin{align}
\mathbb{P}[\mathcal{\bar R}_{x_{\mu_i}} \geq \nu_i]&\leq \mathbb{P}\Big[R_{x_{\mu_i}} \leq \ceil{R^*}\Big]\nonumber \\
&=\int_{1}^{\ceil{R^*}} f(R_{x_{\mu_i}}) \textup{d} R_{x_{\mu_i}}\,,
\end{align}
\vspace{-1.5em}
\bibliographystyle{IEEEtran}\vspace{-1.0em}
\bibliography{IEEEabrv,Literature}

\end{document}